\documentclass[a4paper,11pt]{article}
\pdfoutput=1
\usepackage[utf8]{inputenc}
\usepackage{amsmath}
\usepackage{fullpage}
\usepackage{hyperref}
\usepackage{cleveref}
\usepackage{cite}
\usepackage{graphicx}
\usepackage{subcaption}
\usepackage{placeins}
\usepackage{multirow}
\usepackage{hhline}
\usepackage{rotating}
\usepackage{todonotes}
\usepackage[htt]{hyphenat}

\hypersetup{
  colorlinks=true,
  linkcolor=red,          
  citecolor=blue
}

\begin{document}
\renewcommand*{\thefootnote}{\fnsymbol{footnote}}
\begin{minipage}{\textwidth}
\flushright\footnotesize CoEPP-MN-18-4\\
MCNET-18-23
\end{minipage}\\[12mm]
\noindent\textbf{\LARGE Average Event Properties from LHC to FCC-hh}\\[2ex]
{\large\textbf{Helen Brooks}\footnote{\texttt{Helen.Brooks@monash.edu}} and \textbf{Peter Skands}\\[1ex]
School of Physics \& Astronomy, Monash University, Clayton VIC-3800, Australia}\\[1ex]
\renewcommand*{\thefootnote}{\arabic{footnote}}
\addtocounter{footnote}{-1}
\begin{abstract}
In the context of design studies for future $pp$ colliders, we present a set of predictions for average soft-QCD event properties for $pp$ collisions at $E_\mathrm{CM} = 14$, $27$, and $100$ TeV. The current default Monash 2013 tune of the \textsc{Pythia}~8.2 event generator is used as the baseline for the extrapolations, with uncertainties evaluated via variations of cross-section parametrisations, PDFs, MPI energy-scaling parameters, and colour-reconnection modelling, subject to current LHC constraints. The observables included in the study are total and inelastic cross sections, inelastic average energy and track densities per unit pseudorapidity (inside $|\eta|\le 6$), average track $p_\perp$, and jet cross sections for 50- and 100-GeV anti-$k_T$ jets with $\Delta R=0.4$, using \textsc{aMC@Nlo} in conjunction with \textsc{Pythia}~8 for the latter. 
\end{abstract}
\vspace*{2mm}

\section{Motivation}

The long-term future of physics at the energy frontier is looking increasingly towards designs of $pp$ colliders operating at CM energies of up to 100 TeV (FCC-hh or SppC)~\cite{Arkani-Hamed:2015vfh,Benedikt:2015poa,CEPC-SPPCStudyGroup:2015csa,CEPC-SPPCStudyGroup:2015esa,Mangano:2017tke,Benedikt:2018ofy}. 
In the context of detector design efforts and basic phenomenology studies for such machines, as well as for the 14-TeV LHC and 27-TeV HE-LHC options, a set of predictions (or, at the very least, extrapolations) are required for elementary quantities characterising event properties such as overall rates (total and inelastic cross sections), tracker occupancies (charged-track densities) and total energy depositions ($\mathrm{d}E/\mathrm{d}\eta$). Going beyond predictions for single events, superpositions of minimum-bias events are also used to estimate the dependence of the overall activity on the number of events per bunch crossing (pile-up), which is projected to reach $\mathcal{O}(1000)$ for peak luminosities of $3\times 10^{35}\mathrm{cm}^{-2}\mathrm{s}^{-1}$ targeted at both HE-LHC and FCC-hh~\cite{Benedikt:2018ofy,Benedikt:2018lgy}. 

\begin{figure}[t]
  \centering
  \includegraphics[scale=0.45]{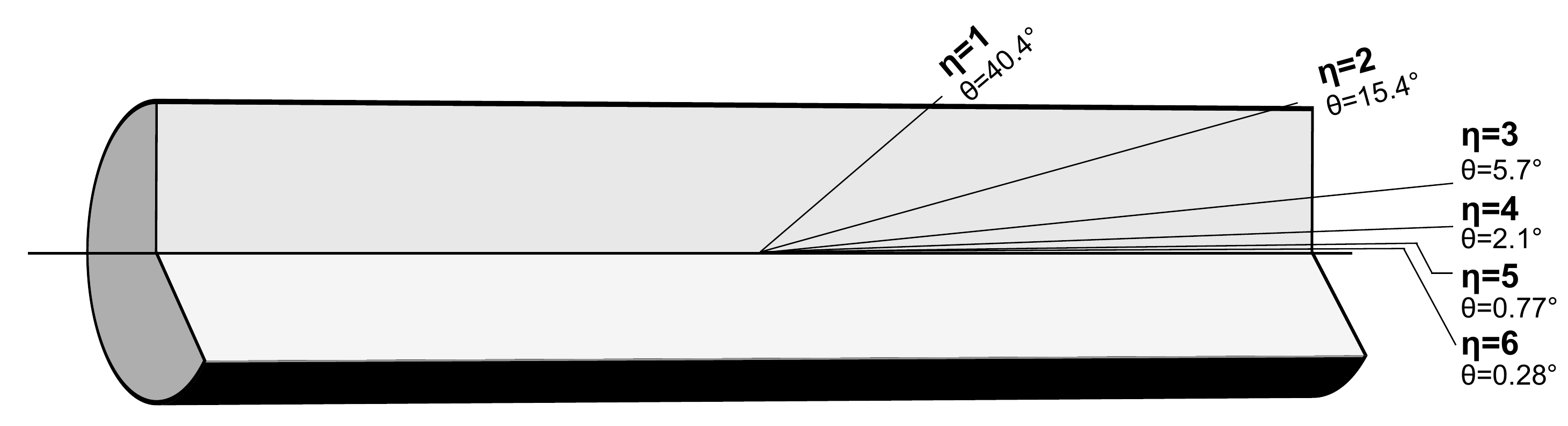}
  \caption{Schematic showing the correspondence between pseudorapidities and angular coverage envisioned 
   for tracking and calorimetry at FCC-hh. (The reference cylinder  indicates the proportions of an instrumented region 
   roughly 55m long with a 6m radius.)  \label{fig:detector}}
\end{figure}

The above are all observables which are dominated by soft QCD effects, and which can therefore not be computed perturbatively. Instead, one is forced to rely either on parametric fits or on explicit physics models such as those implemented in (soft-inclusive) Monte Carlo event generators. In the context of the latter, we here consider the modelling offered by the \textsc{Pythia}~8 event generator~\cite{Sjostrand:2014zea}, and the degree to which its extrapolations are affected by several known sources of theoretical uncertainties. For definiteness, we focus on predictions for an instrumented region corresponding to $|\eta|\le 6$, illustrated in fig.~\ref{fig:detector}. 

After verifying that the default Monash 2013 tune~\cite{Skands:2014pea} of the \textsc{Pythia} 8.2 event generator
still gives an acceptable description of soft-inclusive observables at the LHC including new measurements at 13 TeV, we consider its extrapolations to higher CM energies and perform several salient variations, e.g., of total cross sections~\cite{Rasmussen:2018dgo}, colour reconnections, PDF sets, and multi-parton interactions. 
This study updates and extends the  Snowmass white paper in~\cite{Skands:2013asa} which was based on \textsc{Pythia}~6.4~\cite{Sjostrand:2006za}. It is complementary to the that of \cite{dEnterria:2016oxo} which considered extrapolations of the default predictions (i.e., without parameter variations) of several qualitatively different MC models of soft-inclusive QCD reactions, including \textsc{Epos}~\cite{Pierog:2013ria}, \textsc{Phojet}~\cite{Engel:1995yda}, and \textsc{Qgsjet}~\cite{Ostapchenko:2010vb}. We verify that we obtain consistent results for the common reference model considered in both studies (default \textsc{Pythia} 8.2), and note that most of the models considered in \cite{dEnterria:2016oxo} exhibit a rather similar scaling behaviour of, e.g., the central charged-particle densities over the extrapolated region, predicting that it should grow by about a factor of 2 from 10 to 100 TeV. The exception is the significantly slower scaling exhibited by \textsc{Phojet}~1.12 which is however already in strong conflict  with the LHC measurements, hence we do not consider it a realistic variation\footnote{This is shown more quantitatively in section~\ref{sec:tunevar}. 
Note also that, since \textsc{Pythia} generally predicts higher densities and faster growth than both \textsc{Phojet} and the similar \textsc{Sybill} 2.1 generator~\cite{Ahn:2009wx}, see e.g.~\cite{Skands:2013asa}, the \textsc{Pythia}~8 predictions can also simply be taken as being conservative in this context.}. 

We are thus reasonably confident that the Monash tune provides an acceptable starting point for extrapolations up to 100 TeV, and that useful uncertainty estimates can be obtained by variations of its salient parameters. As a desirable side effect, \textsc{Pythia}'s universal modelling of both hard and soft processes then also allows the same models and variations to be (re)used in the context of studies of hard high-$p_\perp$ processes as well. In the context of such hard (perturbative) processes at 100 TeV, the reader may also be interested in the dedicated study performed in \cite{Bothmann:2016loj}.

The report is organised as follows. 
In section \ref{sec:py8}, we briefly recapitulate the physics of soft QCD processes in the context of \textsc{Pythia}~8, with emphasis on inclusive soft-QCD cross sections and the modelling of MPI and colour reconnections. In section \ref{sec:tunevar}, we summarise the variations that we have performed for uncertainty estimates, along with the LHC constraints imposed to limit the variations. The main set of results for FCC-hh are presented in section \ref{sec:predictions}, with a brief ``executive summary'' of our main conclusions provided in section \ref{sec:conclusions}.

\section{Overview of soft QCD in Pythia 8 \label{sec:py8}}
We shall start by providing a brief review of the modelling of soft QCD processes in
\textsc{Pythia}~8~\cite{Sjostrand:2014zea}, placing particular emphasis on 
those areas relevant to the (re)tuning performed in \cref{sec:tunevar}, in
which variants of the default tune settings are determined. These variants will be 
used to provide uncertainty ranges on predictions extrapolated to FCC-hh energies in \cref{sec:predictions}.

\subsection{Modelling of the total, elastic and diffractive cross sections}
\label{sec:sigma_model}

The total cross section is given by the sum of elastic (el) and inelastic cross sections; the latter is further broken down into the sum of non-diffractive (ND), single diffractive (SD) and  double diffractive 
(DD) cross sections, so that the total cross section is expressed as a sum of four components, 
\begin{equation}
 \sigma_\mathrm{tot}(s) = \sigma_\mathrm{el}(s) + \sigma_\mathrm{ND}(s) + \sigma_\mathrm{SD}(s) + \sigma_\mathrm{DD}(s)~. \label{eq:total}
\end{equation}
The distinction between the three components of the inelastic cross section is in principle slightly model dependent (related to exactly how one distinguishes high-mass diffractive excitations from non-diffractive events), 
but this is beyond the level of accuracy required for this study. We note also that \textsc{Pythia} allows for a small central-diffractive component to be added as well, typically at the level of ${\cal O}(1\%)$ of the total cross section; however this option is switched off by default and has not so far been included in any tunes. Since it is
primarily relevant for dedicated central-exclusive studies, beyond the level of detail aimed for in this study, 
it is neglected in the  following discussion.

In \textsc{Pythia} 8.2 by default the total cross section is modelled by the Donnachie-Landshoff (DL) parameterisation \cite{Donnachie:1992ny}:
\begin{equation}
 \sigma_\mathrm{tot}(s) = X^{AB}s^\varepsilon + Y^{AB}s^{-\eta}
\end{equation}
where $\varepsilon=0.0808$, $\eta=0.4525$, $A$ and $B$ denote the initial state particles and $X^{AB},Y^{AB}$ depend upon these states.
The default elastic and diffractive cross sections are described by the Schuler-Sj\"ostrand (SaS) framework \cite{Schuler:1993wr}. This is a diagrammatical approach based on the exchange of pomerons and reggeons
and application of the optical theorem, similar to the Ingelman-Schlein approach \cite{Ingelman:1984ns}. The non-diffractive cross section, which is calculated from the total, elastic, and diffractive ones by solving for $\sigma_\mathrm{ND}$ in \cref{eq:total}, not only determines the total rate of events that will be modelled as non-diffractive 
(typically starting from a partonic $2\to 2$ QCD interaction process), but is also used for determining the expected number of multi-parton interactions, which we discuss in greater detail in \cref{sec:MPItheory}.

The DL and SaS parameterisations worked well up to Tevatron energies; however the diffractive cross section grows somewhat too quickly with energy and already overshoots the data at LHC energies \cite{ATLAS:2010kza}.
For this reason, some alternatives to these models have now been implemented in \textsc{Pythia} 8. 
First there is the Minimum Bias Rockefeller (MBR) implementation \cite{Ciesielski:2012mc} of a model based on Regge theory \cite{Goulianos:2002vm,Goulianos:2004as}; this primarily addresses diffraction but
also provides parameterisations of the total and elastic cross sections. 
Recently \cite{Rasmussen:2018dgo}, two further options for the total and elastic cross section were implemented: the ABMST model \cite{Appleby:2016ask} and a parameterisation by the COMPAS group found in the Review of Particle Physics 2016 (RPP) \cite{Patrignani:2016xqp}
\footnote{It should be noted that the PDG version contained some misprints: for the formulae that have been implemented in \textsc{Pythia} 8, see \cite{Rasmussen:2018dgo}.}.
The former is an extension of DL, allowing four types of single exchanges, and all combinations thereof for double exchanges; the RPP model is further complicated still, allowing six different single exchanges
and certain combinations of double exchanges.

Additionally, the ABMST model also addresses single diffraction, although it was observed that this also predicts a diffractive cross section which grows faster than the total cross section with energy.
Therefore some modifications to this model giving a more reasonable high energy behaviour were made in the context of the implementation in \textsc{Pythia}~8~\cite{Rasmussen:2018dgo}.


\subsection{Modelling of multi-parton interactions}
\label{sec:MPItheory}

In \textsc{Pythia} 8 the modelling of multi-parton interactions (MPI) \cite{Sjostrand:1987su,Corke:2011yy,Sjostrand:2017cdm} represents an effort to extend perturbative QCD down to low transverse momenta. 
Perturbatively, the hard scattering cross section above some cutoff ${p_\perp}_\mathrm{min}$ is given by
\begin{equation}
 \sigma_\mathrm{hard}({p_\perp}_\mathrm{min}) = \int^{s/4}_{{p^2_\perp}_\mathrm{min}} \frac{\mathrm{d}\sigma}{\mathrm{d}p^2_\perp} \mathrm{d}p^2_\perp.
\end{equation}
 with the differential cross section obtained from the factorisation formula:
\begin{equation}
 \frac{\mathrm{d}\sigma}{\mathrm{d}p^2_\perp}= \sum_{i,j}\int\int \mathrm{d}x_1 \mathrm{d}x_2 f_i(x_1,Q) f_j(x_2,Q) \frac{\mathrm{d}\hat{\sigma}}{\mathrm{d}p^2_\perp}. \label{eq:fact}
\end{equation}
The leading partonic differential cross sections are $t$-channel exchanges which behave as $1/p^4_\perp$ in the $p_\perp \to 0$ limit,  
so $\sigma_\mathrm{hard}$ is divergent for ${p_\perp}_\mathrm{min} \rightarrow 0$ and --- at current LHC energies --- exceeds the total cross section for ${p_\perp}_\mathrm{min} \sim 5$ GeV (see e.g.\ \cite[fig.~34]{Skands:2014pea}). 
This apparent contradiction is reconciled by recognising that $\sigma_\mathrm{hard}$ represents the cross section for not one but \textit{any} number of parton-parton interactions, of which there may be several in a single hadron-hadron collision.

In the limit that the pairwise interactions are independent, the number of parton-parton scatterings per hadron-hadron collision  would be Poisson distributed, with an average given by the ratio of the hard to non-diffractive (ND) cross section\footnote{The fact that the average number of MPI is inversely proportional to $\sigma_\mathrm{ND}$ is easy to understand; 
if we keep $\sigma_\mathrm{hard}$ (i.e., the rate of parton-parton collisions) unchanged, but, say,
reduce $\sigma_\mathrm{ND}$ (expressing the rate of hadron-hadron ones), 
then the average number of parton-parton collisions per hadron-hadron one has to go up.}:
\begin{equation}
 \langle n \rangle({p_\perp}_\mathrm{min}) = \frac{\sigma_\mathrm{hard}({p_\perp}_\mathrm{min})}{\sigma_\mathrm{ND}}~. \label{eq:aveN}
\end{equation}
Within \textsc{Pythia}, this expression is interpreted as the kernel of an ordered sequence of parton-parton interactions~\cite{Sjostrand:1987su}, such that the interaction with the highest $x_\perp=p_\perp/\sqrt{s}$ is generated first. This allows to build in modifications to the PDFs for subsequent MPIs to account for correlations due to momentum and flavour conservation and to interleave the generation of MPI with the shower evolution in a single common sequence of ordered $p_\perp$ values~\cite{Sjostrand:2004ef}. All of these aspects generate deviations from the simple Poissonian ansatz, with momentum conservation being the most important.

While \cref{eq:aveN} resolves the issue of why $\sigma_\mathrm{hard}$ may exceed the total cross section, the former still exhibits a divergence as ${p_\perp}_\mathrm{min} \rightarrow 0$. This is a problem
because the average scattering energy does not decrease with ${p_\perp}_\mathrm{min}$ as quickly as the average number of scatterings rises. The divergence in $\sigma_\mathrm{hard}$ therefore naively implies
that the total scattering energy becomes infinite. However, use of the partonic cross-section assumes free incoming partons, which is not valid for such low transverse momenta where partons become strongly bound.
Instead one can employ a model of effective colour-screening, where colour charges cannot be resolved over distances $d\sim1/p_{\perp 0}$. The 
partonic cross section is thus regularised according to:
\begin{equation}
\frac{\mathrm{d}\hat{\sigma}}{\mathrm{d}p^2_\perp}\propto  \frac{\alpha_s(p^2_\perp)}{p^4_\perp}\rightarrow \frac{\alpha_s(p^2_{\perp 0}+p^2_\perp)}{(p^2_{\perp 0}+p^2_\perp)^2}
\end{equation}
and the effective screening scale $p_{\perp 0}$ is given  an energy dependence:
\begin{equation}
p_{\perp 0}(E_\mathrm{CM}) =p^\mathrm{ref}_{\perp 0}\left(\frac{E_\mathrm{CM}}{E^\mathrm{ref}_\mathrm{CM}}\right)^{E^\mathrm{pow}_\mathrm{CM}} \label{eq:pt0}
\end{equation}
The arbitrary  reference energy $E^\mathrm{ref}_\mathrm{CM}$ is typically set to 7 TeV in modern tunes, while the parameters $p^\mathrm{ref}_{\perp 0}$ and $E^\mathrm{pow}_\mathrm{CM}$ should be tuned. Essentially, $p^\mathrm{ref}_{\perp 0}$ still sets the average number of MPI per (inelastic, non-diffractive) hadron-hadron collision at the given reference CM energy (for given  $\sigma_\mathrm{ND}$ and $\sigma_\mathrm{hard}$ parameters), 
while  $E^\mathrm{pow}_\mathrm{CM}$ controls the scaling of that number to other CM energies, 
with lower values of $E^\mathrm{pow}_\mathrm{CM}$ producing faster scalings of $\left<n\right>$.

The number of scatterings should also reflect the fact that the incoming hadrons are extended objects, with more peripheral collisions expected to have fewer scatterings on average.
This can be quantified through the impact parameter $b$; although the matter profile of the hadrons is not known a priori, one option\footnote{Other options, including a double Gaussian profile, are also available
in \textsc{Pythia} \cite{Sjostrand:1987su,Corke:2011yy}.} is to parameterise the time-integrated 
overlap function directly
\begin{equation}
 \mathcal{O}(b) \propto \exp(-b^p) \label{eq:overlap}
\end{equation}
where the power $p$ is a free parameter to be tuned, with e.g.\ $p=2$ corresponding to a Gaussian density profile. 
Empirically, values between 1.5 and 2 are preferred. The average number of interactions (per event having at least one scattering) at impact parameter $b$ is given by
\begin{equation}
 \langle n (b)\rangle = \frac{ k \mathcal{O}(b)}{1 - \exp(-k\mathcal{O}(b))}
\end{equation}
and $k$ is a constant to be found by integrating over $b$ and comparing to \cref{eq:aveN}. This may be used to find a probability distribution dependent on both $x_\perp$ and $b$. Thus, when the first (hardest) MPI is generated, an overall $b$ is selected as well, and the corresponding overlap factor is used to enhance (for smaller-than-average $b$) or suppress (for larger-than-average $b$) the probability for all subsequent MPI as well.  

\subsection{Modelling of colour reconnections}
\label{sec:CR}

The Lund model of hadronisation implemented in \textsc{Pythia} interprets the 
linear behaviour of the quark-antiquark potential at long distances as a string which can fragment into hadrons through quantum mechanical tunnelling; see e.g.~\cite{Andersson:1983ia}.
In particular, for a given event the strings are identified as occurring between colour-connected partons, with quarks (and antiquarks) as string endpoints and gluons appearing as transverse ``kinks'' on these strings. 
The picture of colour connections is particularly simple in the leading colour approximation used in nearly all current parton showers\footnote{There have been attempts to go beyond the leading colour approximation in 
the parton shower \cite{Platzer:2012np,Nagy:2012bt,Nagy:2015hwa,Martinez:2018ffw,Isaacson:2018zdi,Platzer:2018pmd},
however these are not formulated as full-fledged event generators, and therefore cannot incorporate such non-perturbative effects as MPI and colour reconnections, and must project back to leading colour prior to hadronisation.}, where each successive emission generated is colour connected to its parent emitters.

Models of colour reconnection allow for strings to form between partons beyond these simple leading colour topologies
effectively allowing for either interference between different flows (`static' colour reconnection) or soft gluon exchanges (`dynamic' colour reconnection). Such effects were already seen to play a role at LEP in measurements of the $W$ boson mass \cite{Schael:2013ita}; 
they become even more important at the LHC due to the existence of vastly more possibilities for coloured initial state partons, beam remnants and multi-parton interactions, all allowing for non-trivial (reconnections of) coloured topologies. 
The leading assumption is that colour reconnection models should give rise to a greater number of shorter strings, thereby affecting the number and energy density distributions of final state particles. 
There are by now many indications that such effects are essential for modelling soft QCD effects in general, 
and there are important open questions concerning their impact on precision observables such as the hadronic top quark mass~\cite{Skands:2007zg,Argyropoulos:2014zoa}.

In \textsc{Pythia} 8 several models for colour reconnections have been developed and implemented over the years, however it remains 
one of the least well understood phenomena in hadron collisions, and in particular we know little about how such effects scale with energy \cite{Schulz:2011qy}. 
Therefore in assessing uncertainties on predictions for future colliders,
it is important not just to include variations on the baseline model for colour reconnections, but also consider some qualitatively different models which may exhibit a different scaling. 

In the oldest model (used by default), gluons from MPI with low transverse momentum scales may be successively inserted into the colour flow of a higher transverse momentum MPI, 
in such that the total string length is minimised by brute force.
A more recent model has also been implemented \cite{Christiansen:2015yqa}, which combines the earlier string-length minimisation arguments with selection rules based
on the colour algebra of SU(3).
Another option known as the `gluon-move' model \cite{Argyropoulos:2014zoa} is also available, which we mention here for completeness; however in  the following section we only consider tunes using the former two.
In this model, any pair of gluons can be reconnected, irrespective of whether they were produced in the same or different MPI. A string-length-reduction measure is calculated,
and pairs of gluons maximising this measure (or minimising change in string length) are sequentially reconnected until the incremental reduction in string-length is below some threshold.


\section{Variants of Monash 2013 Tune and Validation}
\label{sec:tunevar}

In the next section, we present predictions for FCC-hh of certain distributions sensitive to soft QCD, using the Monash 2013 tune of \textsc{Pythia} 8.2 as the baseline. 
Despite being a few years old by now, this tune remains a useful benchmark; as we show below, it remains in acceptable agreement with soft-inclusive LHC measurements up to 13 TeV, and as the default tune for \textsc{Pythia}~8.2 there is a large body of existing validations and further complementary studies can be done without requiring any special parameter settings. 
Asides from being a convenient choice, we also estimate that its predictions for future energies are likely to be a bit on the conservative side, since already at 13 TeV it appears to slightly overestimate the number density of charged particles (see section \cref{sec:tunePDF} and \cref{fig:NchIncrease}). 

Since there are substantial uncertainties arising from ambiguities in the modelling of non-perturbative physics, however, we do not restrict ourselves to a single tune. Instead, in this section we explore variants of the baseline tune in which the parameters associated with the  modelling aspects discussed in the previous section (modelling of PDFs, MPI, total, elastic and diffractive cross sections, and colour reconnections), are modified, one by one. We shall then use the baseline tune together with these variants to define an uncertainty envelope for the extrapolations to higher CM energies in \cref{sec:predictions}. 

\subsection{PDF variation}
\label{sec:tunePDF}

\begin{figure}[p]
\centering
 \begin{subfigure}{\textwidth}
 \centering
\includegraphics[width=\textwidth]{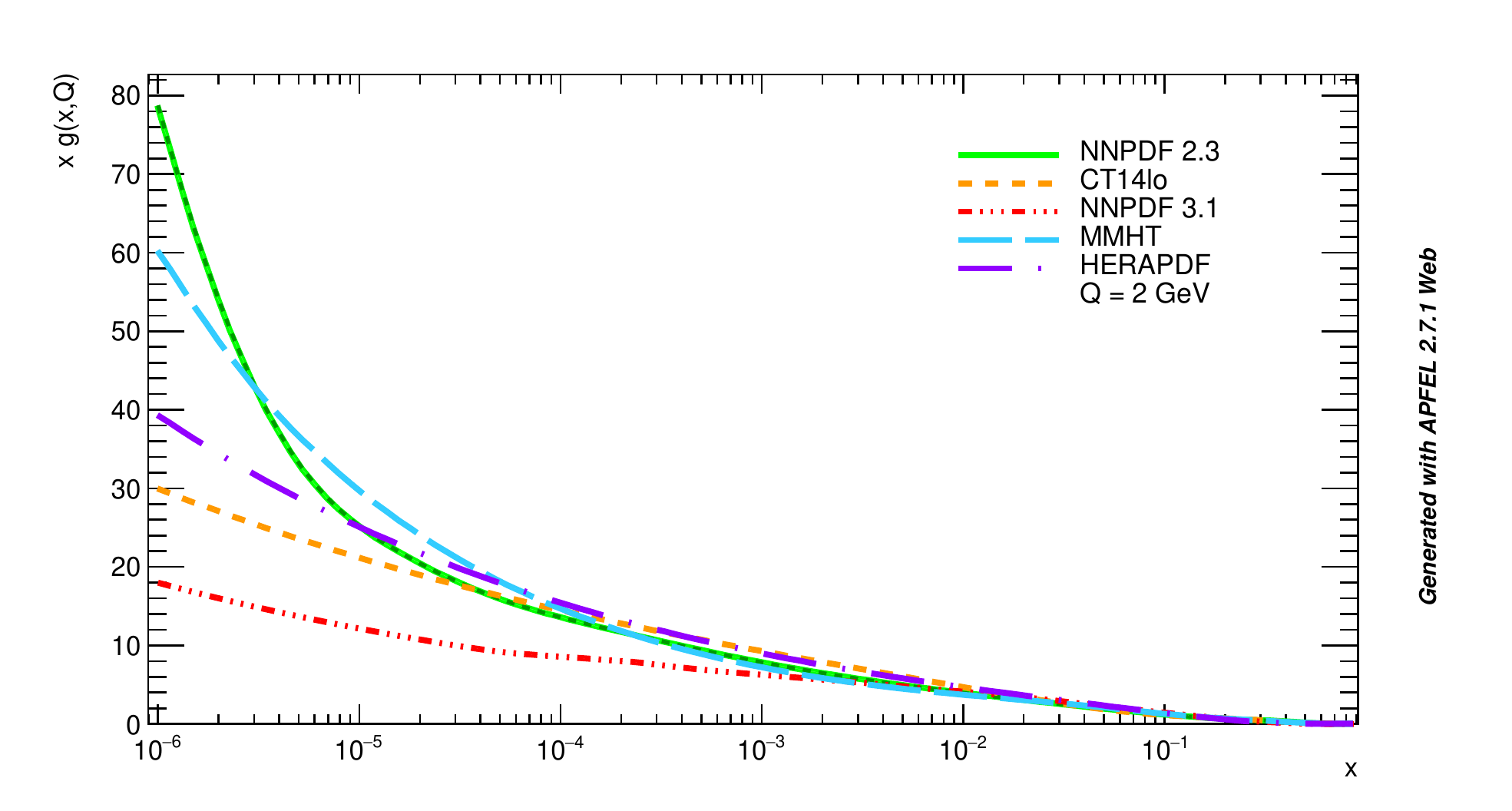}
 \caption{Comparison of LO PDF sets, central predictions only. The central member of the NNPDF 2.3 set used in the default Monash 2013 tune exhibits the most steeply rising gluon distribution at low $x$. The central member of the NNPDF 3.1 set represents a drastic change, exhibiting the slowest rise of all the sets considered here. Among the more recent sets (i.e., discounting NNPDF 2.3), the MMHT set exhibits the fastest rise, while HERAPDF and CT14lo are intermediate between MMHT and NNPDF 3.1.}
 \end{subfigure}
 
  \begin{subfigure}{\textwidth}
 \centering
\includegraphics[width=\textwidth]{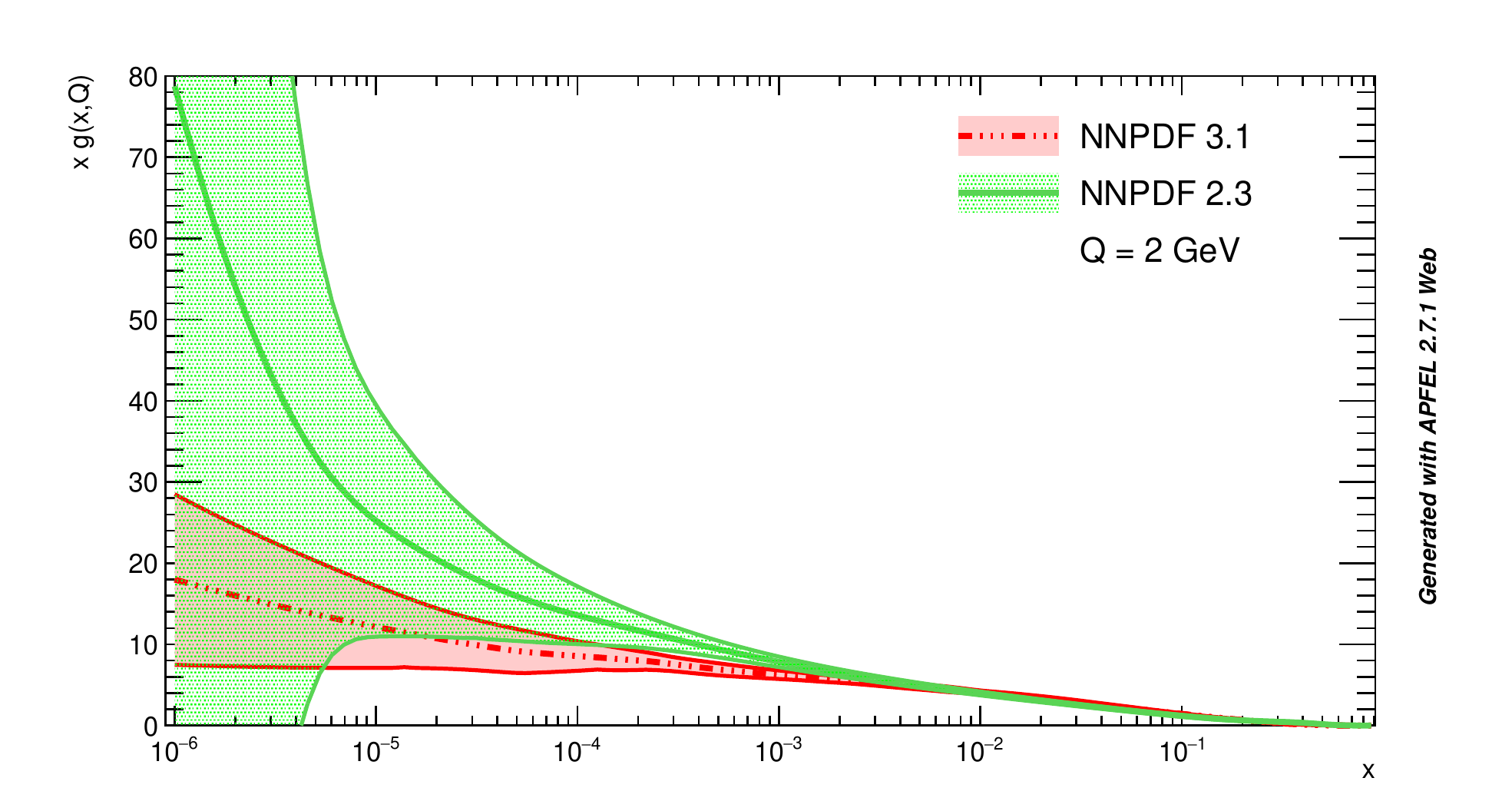}
 \caption{Comparison of NNPDF 2.3 and 3.1 PDF sets showing 1$\sigma$ standard deviation. }
 \end{subfigure}
 
 \caption{A comparison of LO gluon distributions, $xg(x,Q=2\ \mathrm{GeV})$. Plots generated using the APFEL web interface \cite{Bertone:2013vaa,Carrazza:2014gfa}).}
 \label{fig:comparegPDF}
\end{figure}

We recall from  \cref{sec:MPItheory} that the probability distribution for multiple scatterings is dependent on $\mathrm{d}\sigma/\mathrm{d}p^2_\perp$ and hence on the PDFs through \cref{eq:fact}.
MPI are primarily driven by the low-$x$ behaviour of the gluon PDF, which is not tightly constrained and can vary significantly between different sets. In addition to the baseline NNPDF 2.3 QCD+QED LO set (with $\alpha_s(m_Z)=0.13$), the following three LO PDF sets (interfaced  via LHAPDF~6~\cite{Buckley:2014ana}) were chosen as spanning a reasonable range of qualitatively different behaviours:
\begin{itemize}
 \item NNPDF31\_LO\_as\_0130 \cite{Ball:2017nwa}
 \item HERAPDF15LO\_EIG \cite{CooperSarkar:2011aa}
 \item MMHT2014lo68cl \cite{Harland-Lang:2014zoa}
\end{itemize}
In \cref{fig:comparegPDF}, we show a comparison of the gluon distributions of these sets at a relatively low scale, $Q=2\ \mathrm{GeV}$, representative of a ``typical'' scale probed by MPI. Note that the CT14lo \cite{Dulat:2015mca} set is also shown\footnote{CT14llo was also considered but is not shown as it was not significantly different from CT14lo.} but was not found to add qualitatively to the range already spanned by the other sets.

To define Monash tune variants using the alternative PDF sets, we re-tune a few of the parameters which control the amount of MPI to reobtain a basic level of agreement with LHC measurements; see below. At the very least, we consider the $p_{\perp 0}$ parameter discussed in \cref{sec:MPItheory} to be intimately tied to the choice of PDF and hence obligatory to re-tune when the PDF is modified. For slightly more flexibility, we allow the following three parameters to be re-tuned when changing the PDF set:
\begin{itemize}
 \item \texttt{MultipartonInteractions:pT0Ref}
 \item \texttt{MultipartonInteractions:ecmPow}
 \item \texttt{MultipartonInteractions:expPow}
\end{itemize}
which control $p^\mathrm{ref}_{\perp 0}$ and $E^\mathrm{pow}_\mathrm{CM}$ in \cref{eq:pt0}  and $p$ in \cref{eq:overlap} respectively.

\begin{figure}[p]
\centering
 \begin{subfigure}{0.45\textwidth}
 \centering
\includegraphics[width=\textwidth]{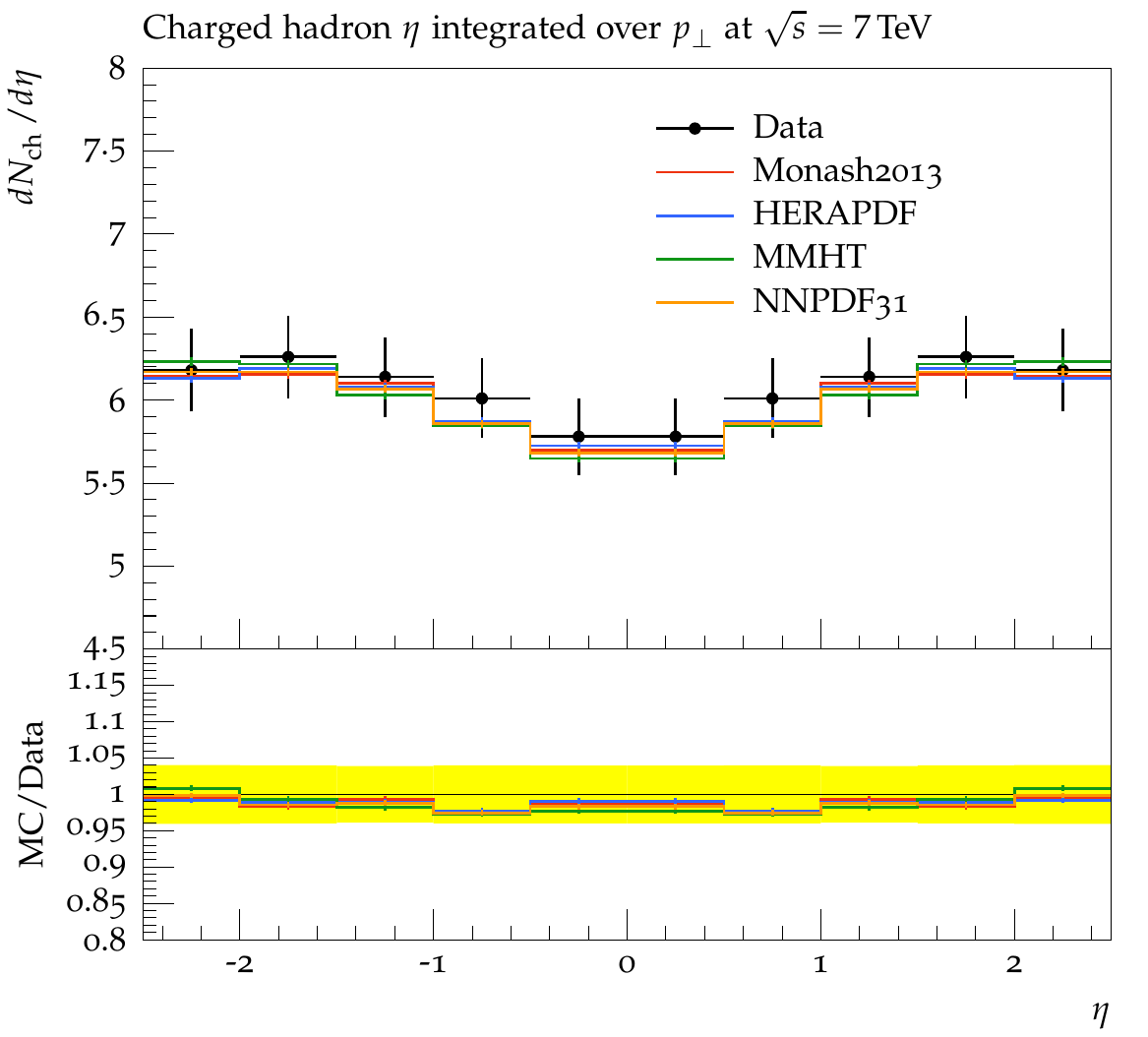}
\caption{}
\label{fig:tunepT0}
 \end{subfigure}
 \begin{subfigure}{0.45\textwidth}
 \centering
\includegraphics[width=\textwidth]{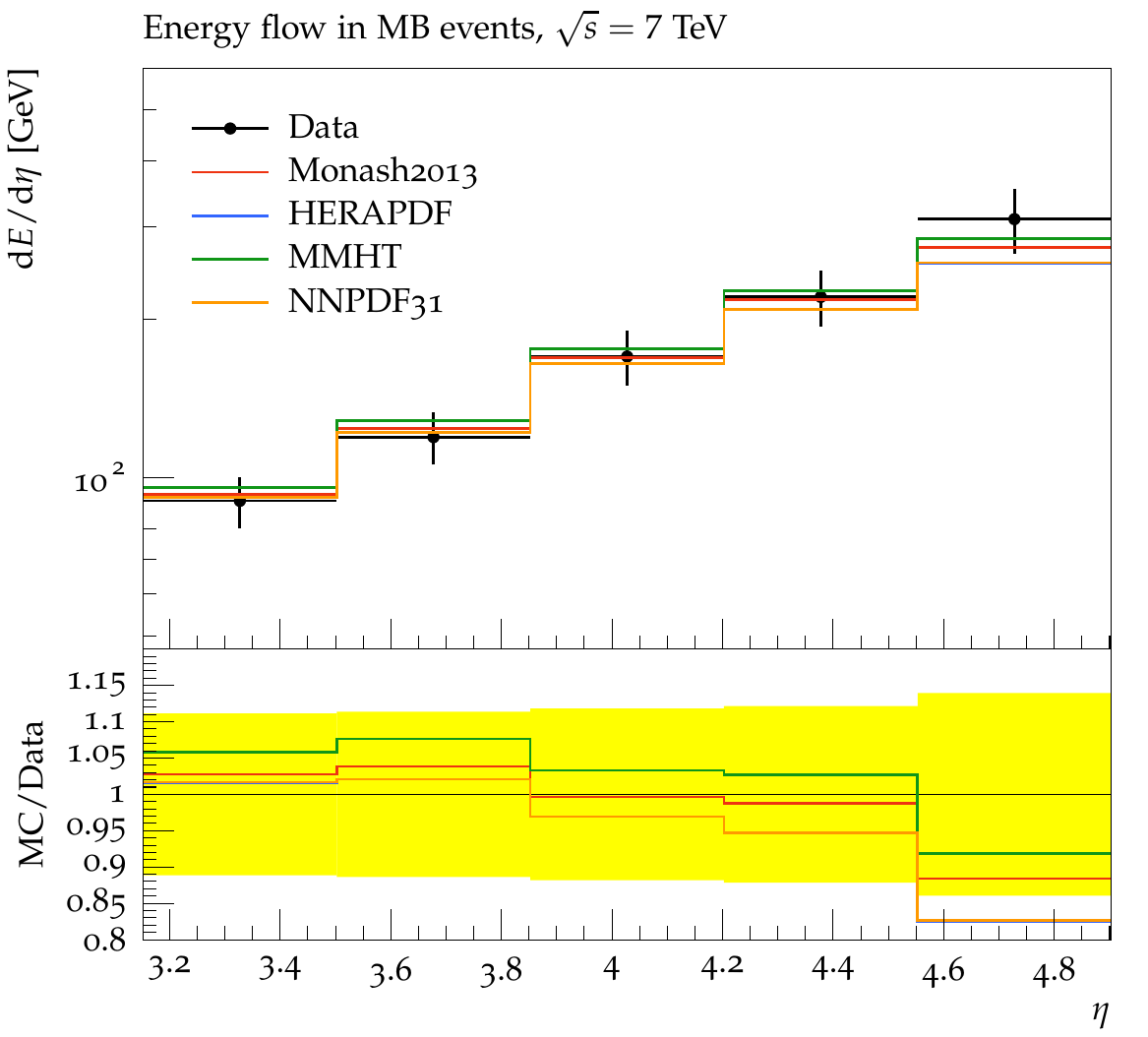}
\caption{}
\label{fig:forwardE}
\end{subfigure}
 \begin{subfigure}{0.45\textwidth}
 \centering
\includegraphics[width=\textwidth]{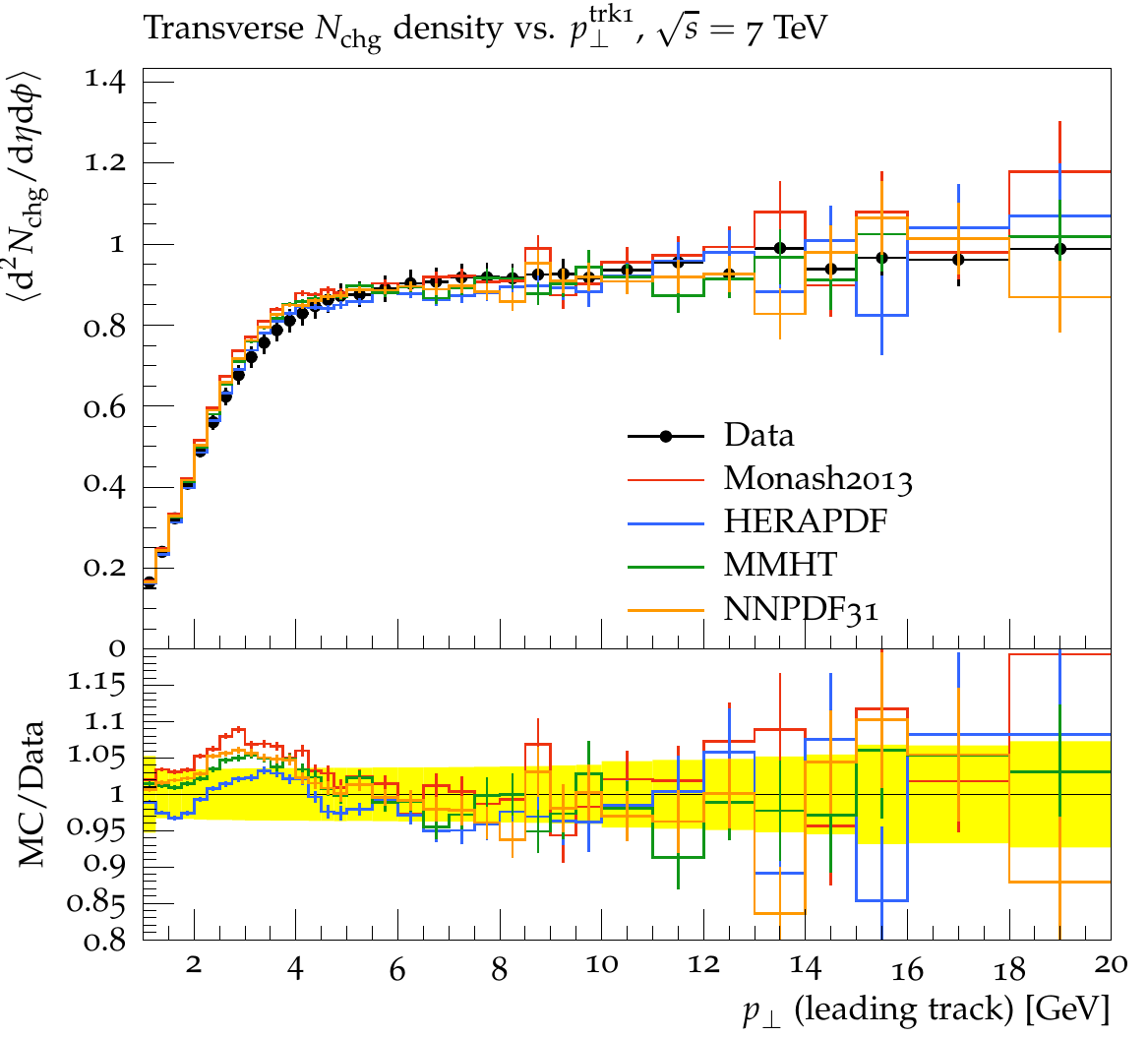}
 \caption{}
 \label{fig:nchdens_atlas}
 \end{subfigure}
 \begin{subfigure}{0.45\textwidth}
 \centering
\includegraphics[width=\textwidth]{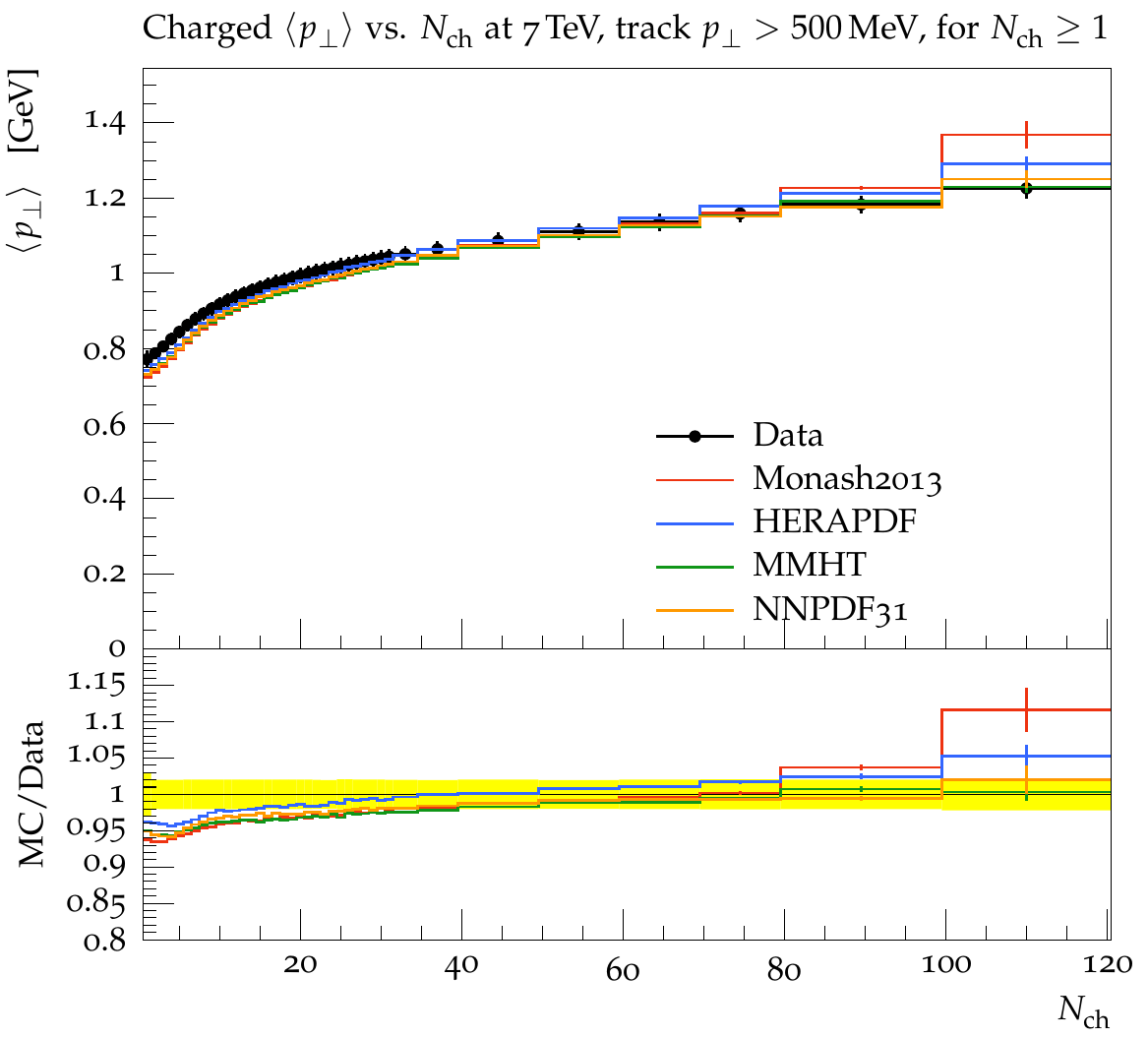}
\caption{}
\label{fig:avept_atlas}
\end{subfigure}
 \caption{A selection of distributions measuring the underlying event level at the LHC at 7 TeV which were used in the tuning of variants of Monash 2013 with different PDF selection, described in the text. 
 The differential number of charged tracks in pseudorapidity as measured by CMS \cite{Khachatryan:2010us} is shown in \subref{fig:tunepT0}.
 The differential energy deposit at large rapidities as measured by CMS \cite{Chatrchyan:2011wm} is shown in \subref{fig:forwardE}.
 The density of charged tracks as a function of the transverse momentum of the hardest track as measured by \cite{Aad:2010fh} is shown in \subref{fig:nchdens_atlas}.
 The average transverse momentum of charged tracks as a function of the number of charged tracks measured by ATLAS \cite{Aad:2010ac} is shown in  \subref{fig:avept_atlas}.
 Plots were produced using Rivet \cite{Buckley:2010ar}.
 }
 \label{fig:tunePDFs}
\end{figure}

For the tuning, a variety of analyses measuring the underlying event and minimum bias at the LHC at 7 TeV, implemented in Rivet \cite{Buckley:2010ar}  were considered.
The value of \texttt{MultipartonInteractions:pT0Ref} was tuned such that the number of charged tracks in $\eta$, $\mathrm{d}N_\mathrm{ch}/\mathrm{d}\eta$
was compatible with the level measured, for example, by CMS \cite{Khachatryan:2010us}, as shown in \cref{fig:tunepT0}. The energy flow at high $\eta$~\cite{Chatrchyan:2011wm} is also shown, in \cref{fig:forwardE}, since that is one of the observables we will be considering in the extrapolations to 100 TeV.
The value for \texttt{MultipartonInteractions:expPow} was tuned to the density of charged tracks as a function of the transverse momentum of the hardest track, for example,
as measured by ATLAS  \cite{Aad:2010fh}, cf.~\cref{fig:nchdens_atlas}. Although shape differences arise in this observable, these are no more significant than for the default Monash tune. Finally, we also include a comparison to the average track $p_\perp$ as function of the track multiplicity, in \cref{fig:avept_atlas}, again as a reference for one of the observables we shall consider in our extrapolations to higher energies.

We briefly note that for the setup in \cref{fig:avept_atlas}, namely minimum bias events having at least one charged track with transverse momentum exceeding 500 MeV, the tunes all undershoot the data by about 5\%. It is known that the transverse momentum of charge tracks is imperfectly modelled: see for example figure 18 in \cite{Skands:2014pea}, where the ratio to data of the differential multiplicity with respect to transverse momentum is not flat. This is likely due to an incorrect distribution in the transverse momentum kicks that result from string-breaking during hadronisation (currently modelled by a Gaussian). The result is that a slightly different selection at trigger level can result in a different normalisation relative to data in the average transverse momentum (see for example figure 20 in \cite{Skands:2014pea}). To improve upon this situation will require improvements in the modelling of hadronisation, which we do not discuss further here.

\begin{figure}[t]
 \centering
  \begin{subfigure}{0.45\textwidth}
 \centering
\includegraphics[width=\textwidth]{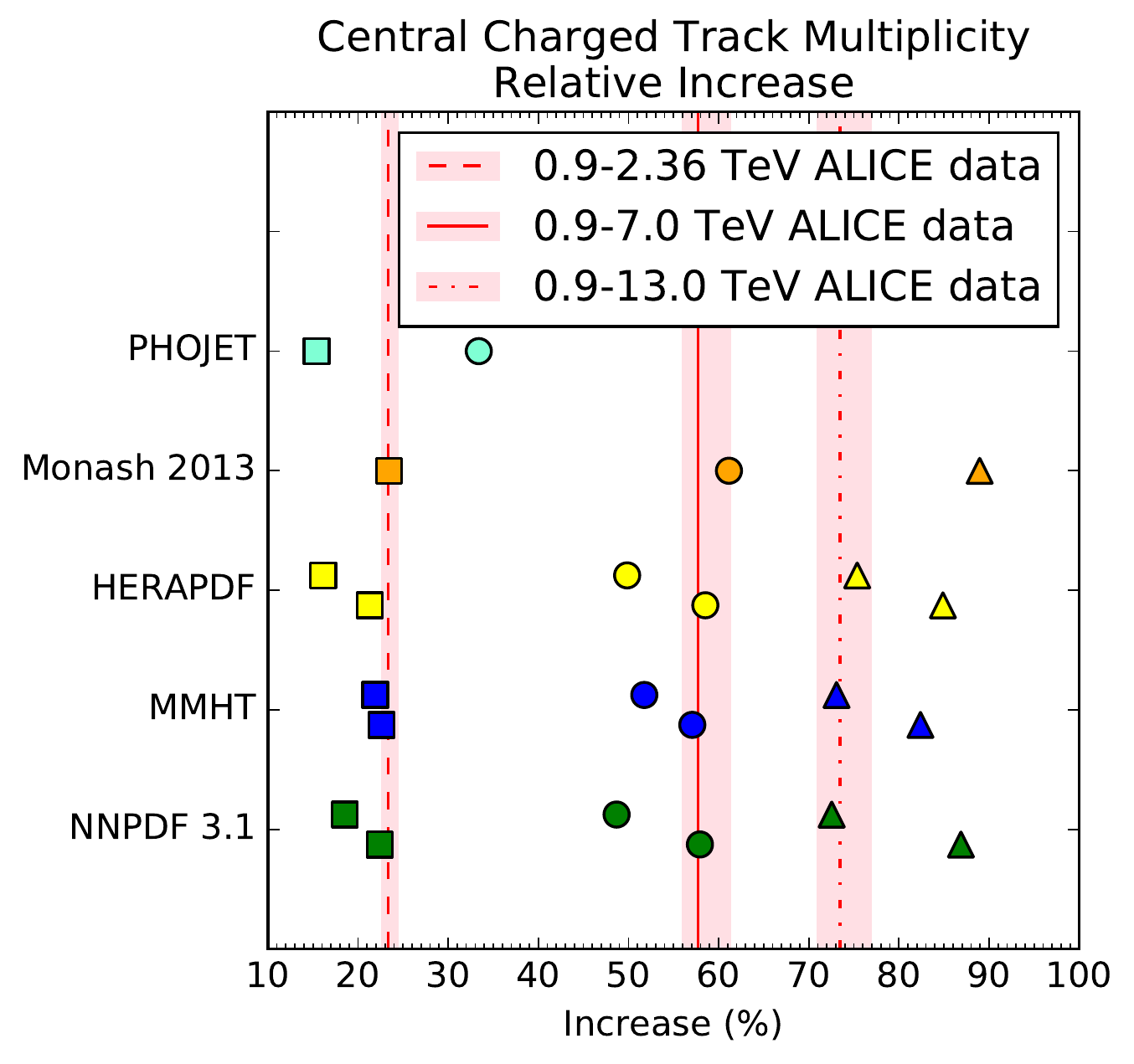}
\caption{PDF Variations (+ PHOJET)}
\label{fig:Nch_PDFS}
 \end{subfigure}
   \begin{subfigure}{0.45\textwidth}
 \centering
\includegraphics[width=\textwidth]{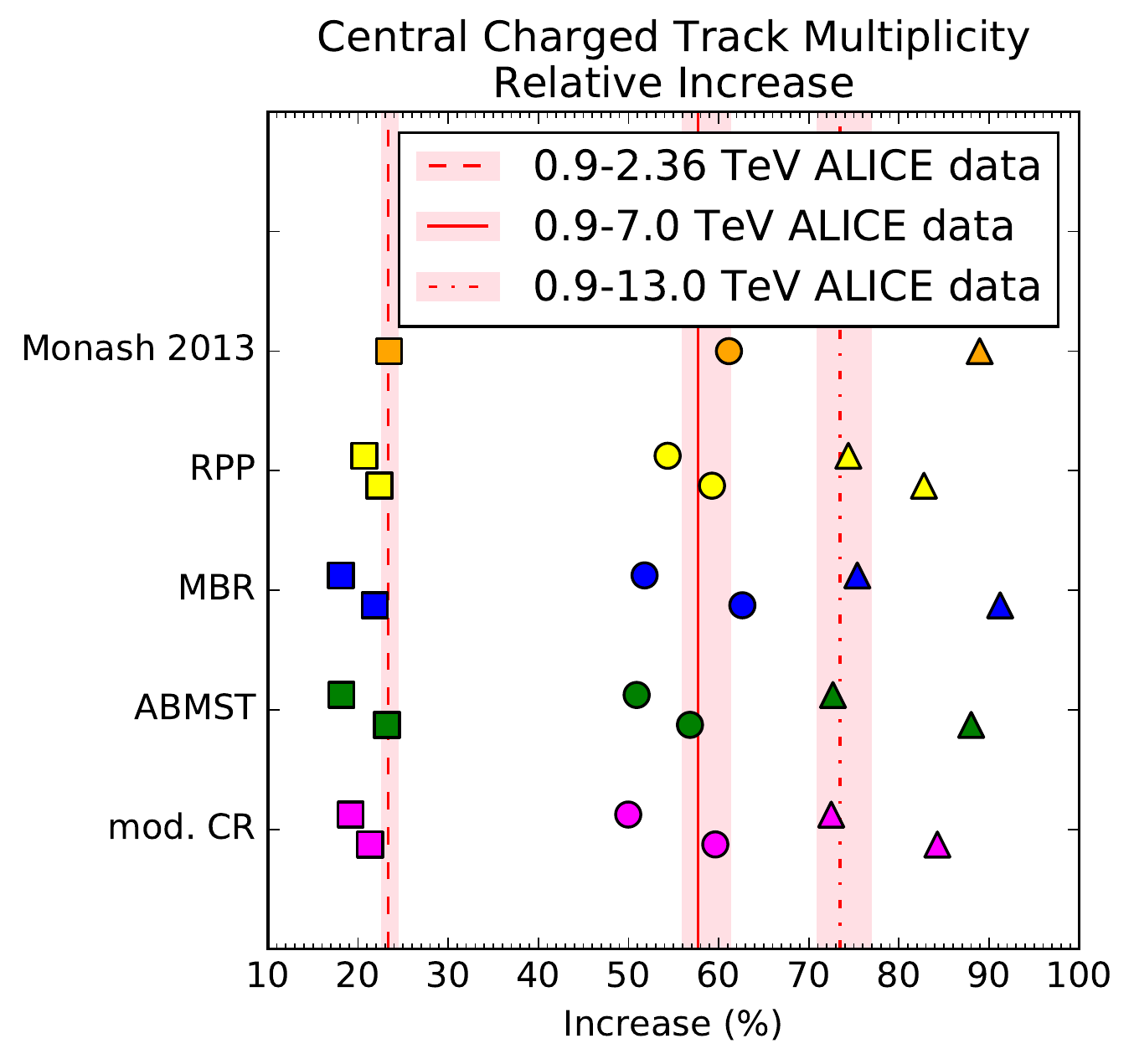}
\caption{Cross-Section and CR Variations}
\label{fig:Nch_sigma}
\end{subfigure}
 \caption{The relative increase in the number density of charged tracks between 0.9-2.36 TeV and 0.9-7 TeV measured by ALICE \cite{Aamodt:2010pp} and updated to include 13 TeV data \cite{Adam:2015pza}.
 This observable was used to tune to \texttt{MultipartonInteractions:ecmPow} for variants of the Monash 2013 tune with \subref{fig:Nch_PDFS} alternative choice for the PDF, and \subref{fig:Nch_sigma}
 different choices for the modelling of the total cross section and colour reconnections. For comparison, the predictions for PHOJET \cite{PhysRevD.52.1459} originally included in \cite{Aamodt:2010pp}
 are also shown in \subref{fig:Nch_PDFS}. In all cases, where two points are shown, these reflect two choices for the parameter \texttt{MultipartonInteractions:ecmPow}, 
 with the lower point corresponding to the minimum value and the upper point corresponding to the maximum value.
 }
 \label{fig:NchIncrease}
\end{figure}

\texttt{MultipartonInteractions:ecmPow}  controls the energy scaling of the $p_{\perp 0}$ regularisation scale for MPI, as discussed in \cref{sec:MPItheory}. This was tuned by considering a particularly sensitive observable,
namely the relative increase in the density of tracks between 0.9-2.36 TeV and 0.9-7 TeV measured by ALICE \cite{Aamodt:2010pp}. We here extend this plot to include recent 13 TeV data 
\cite{Adam:2015pza}, though without the benefit of internal ALICE systematics studies, we cannot take correlations between the 900-GeV and 13-TeV uncertainties into account and merely combine them using simple quadratures, yielding a conservative overestimate which is fine for our purpose. While it is quite easy to tune \texttt{MultipartonInteractions:ecmPow} to capture the relative increase between from 0.9 to 2.36 and 7 TeV, such tunes typically exhibit some tension with the 13 TeV data. Therefore, for each of the changed PDF sets we determine two alternatives - a minimum and maximum value - for this parameter. The results are shown in \cref{fig:Nch_PDFS}, with two points shown at each energy for each of the HERAPDF, MMHT, and NNPDF 3.1 variants; the lower (higher) point corresponding to the minimum (maximum) value for \texttt{MultipartonInteractions:ecmPow}.

For completeness, the plot in \cref{fig:Nch_PDFS} also shows the predictions for PHOJET \cite{PhysRevD.52.1459}, as these were also originally included in \cite{Aamodt:2010pp}. These predictions clearly do not capture the rise in the charged track density, a trend which we expect to be exacerbated when going to 100 TeV. We therefore discount PHOJET as a realistic variation. The final selection of tuned parameters is shown in \cref{table:PDFtunes}.

\begin{table}[h]
\centering
\begin{tabular}{|c|c|c c|c c|c c|}
\hline 
\multirow{2}{*}{Parameter}  & \multirow{2}{*}{default} & \multicolumn{2}{|c|}{NNPDF 3.1}   & \multicolumn{2}{|c|}{HERAPDF} & \multicolumn{2}{|c|}{MMHT} \\
\hhline{~~------}
& & min & max & min & max & min & max\\
\hline 
 \texttt{MultipartonInteractions:pT0Ref} & 2.28 & \multicolumn{2}{c|}{2.22} & \multicolumn{2}{c|}{2.56} & \multicolumn{2}{c|}{2.28} \\
 \texttt{MultipartonInteractions:expPow} & 1.85 & \multicolumn{2}{c|}{1.85} & \multicolumn{2}{c|}{1.72} & \multicolumn{2}{c|}{1.67} \\
 \texttt{MultipartonInteractions:ecmPow} & 0.215 & 0.140 & 0.170 & 0.228 &  0.250 & 0.236 & 0.260 \\
\hline 
\end{tabular}
\caption{Table showing the values of changed parameters relative to the default Monash2013 tune of \textsc{Pythia} 8.2 in variants where the PDF has been changed.}
\label{table:PDFtunes}
\end{table}

\subsection{Variation of cross-section parameterisations}
\label{sec:sigmavartune}
As discussed in \cref{sec:sigma_model} the default parameterisations of the total, elastic and diffractive cross sections in \textsc{Pythia} 8 (namely the DL/SaS parameterisations)
are not able to faithfully reproduce all aspects of the energy scalings seen in measurements, and a set of alternatives were recently implemented~\cite{Rasmussen:2018dgo}. 
For the total cross section we consider the  MBR, ABMST and RPP models (selected by 
modifying \texttt{SigmaTotal:mode}). In the case of the former two, the same models also offer alternative modelling of the diffractive cross section (selected by modifying \texttt{SigmaDiffractive:mode}). Note that in the case of the choice of ABMST for modelling the diffractive cross section, the modifications 
proposed in \cite{Rasmussen:2018dgo} are selected (by using the default value for \texttt{SigmaDiffractive:ABMSTmodeSD}).

\begin{table}[t]
\centering
\begin{tabular}{|c|c|c c|c c|c c|}
\hline 
\multirow{2}{*}{Parameter}  & \multirow{2}{*}{default} & \multicolumn{2}{|c|}{MBR}   & \multicolumn{2}{|c|}{ABMST} & \multicolumn{2}{|c|}{RPP/SaS} \\
\hhline{~~------}
& & min & max & min & max & min & max\\
\hline 
\texttt{SigmaTotal:mode} & 1 & \multicolumn{2}{c|}{2} & \multicolumn{2}{c|}{3} & \multicolumn{2}{c|}{4} \\
\texttt{SigmaDiffractive:mode}& 1 & \multicolumn{2}{c|}{2} & \multicolumn{2}{c|}{3} & \multicolumn{2}{c|}{1} \\ 
 \texttt{MultipartonInteractions:ecmPow} & 0.215 & 0.230 & 0.250 & 0.230 & 0.250 & 0.220 & 0.240 \\
\hline 
\end{tabular}
\caption{Table showing the values of changed parameters relative to the default Monash2013 tune of \textsc{Pythia} 8.2 in variants where the modelling of the total, elastic and diffractive
cross sections has been changed.}
\label{table:sigmatunes}
\end{table}

We only found it necessary to retune one parameter, the energy-scaling parameter \texttt{MultipartonInteractions:ecmPow}, for the above alternative parameterisations of the total/elastic/diffractive cross sections. As in the previous section it was found that the 13 TeV data highlight tensions in the predictions, and accordingly we selected two distinct values instead of a single ``average'' value, as shown in \cref{table:sigmatunes}. The effect of these variations upon the relative increase in the density of charged tracks is shown in \cref{fig:Nch_sigma}.

\subsection{Variation of modelling of colour reconnections}
\label{sec:crtune}

To gauge the uncertainty arising from the still poorly understood issue of colour reconnections, both the default model (based on string minimisation alone) and 
a more recent model 
(based on QCD selection rules) were considered, see the discussion in \cref{sec:CR}. For the latter, the modifications to tuned parameters 
proposed in \cite{Christiansen:2015yqa}
were found to be sufficient. In addition, and as in previous sections, two values for \texttt{MultipartonInteractions:ecmPow} were determined, the effect of which is shown in 
\cref{fig:Nch_sigma}. The full set of changed parameters for these variants is shown in \cref{table:crtune}.

\begin{table}[h]
\centering
\begin{tabular}{|c|c|c c|}
\hline 
\multirow{2}{*}{Parameter}  & \multirow{2}{*}{default} & \multicolumn{2}{|c|}{mod CR} \\
\hhline{~~--}
& & min & max \\
\hline 
\texttt{ColourReconnection:mode} & 0 &\multicolumn{2}{c|}{1} \\
\texttt{BeamRemnants:remnantMode} & 0 &\multicolumn{2}{c|}{1} \\
\texttt{ColourReconnection:allowDoubleJunRem} & on &\multicolumn{2}{c|}{off}\\
\texttt{MultipartonInteractions:pT0Ref} & 2.28 &\multicolumn{2}{c|}{2.15} \\
\texttt{MultipartonInteractions:ecmPow} & 0.215 & 0.215 &0.240\\
\hline 
\end{tabular}
\caption{Table showing the values of changed parameters relative to the default Monash2013 tune of \textsc{Pythia} 8.2 in variants where the modelling of colour reconnections 
has been changed.}
\label{table:crtune}
\end{table}

\clearpage

\section{Predictions for FCC-hh}
\label{sec:predictions}
In this section we present predictions for FCC-hh obtained using the default Monash 2013 tune of \textsc{Pythia} 8 and the variants of this tune described in \cref{sec:tunevar}.

\begin{figure}[tp]
 \centering
\begin{subfigure}{0.45\textwidth}
 \centering
\includegraphics[width=\textwidth]{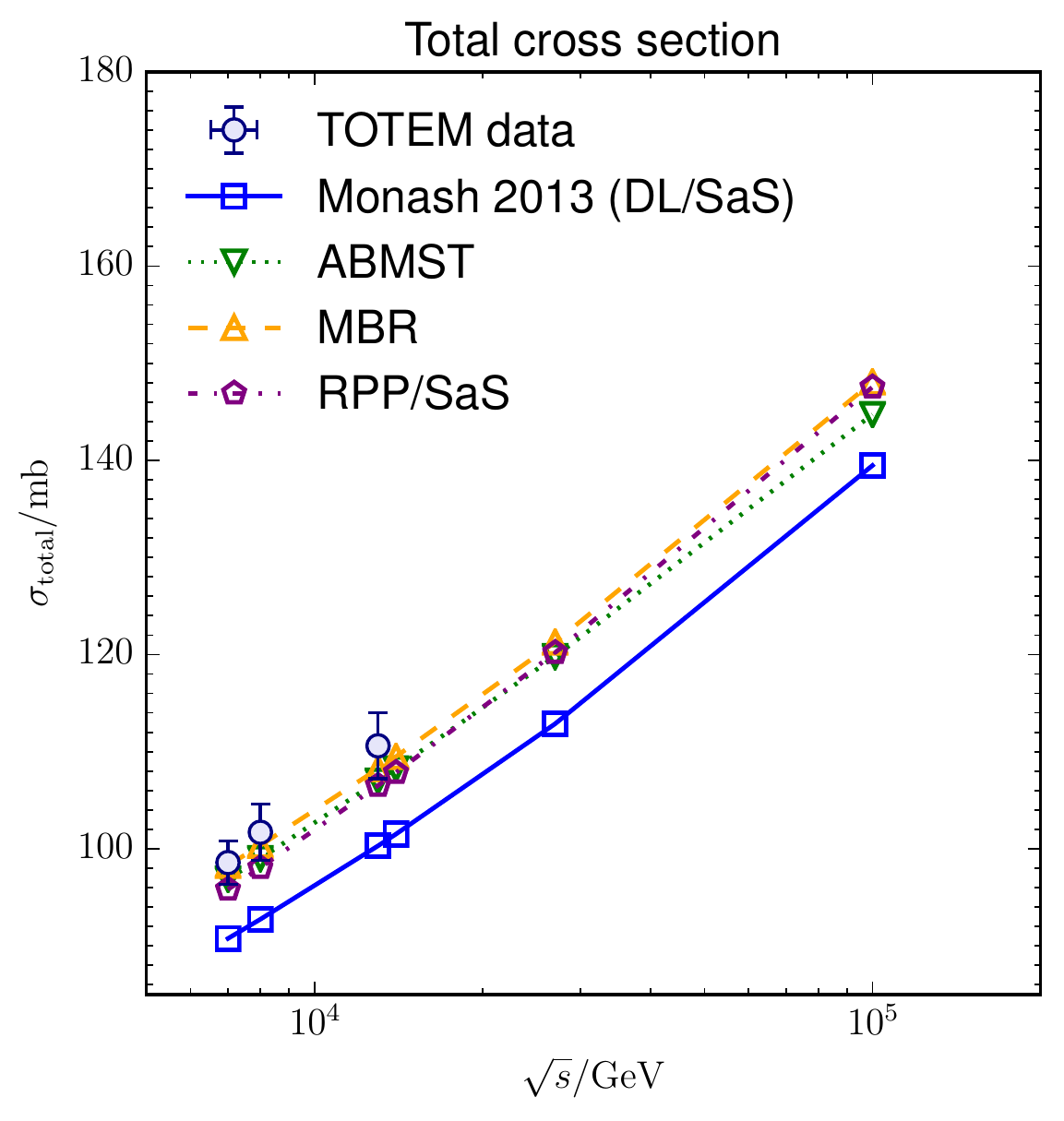}
\caption{}
\label{fig:sigma_total}
 \end{subfigure}
\begin{subfigure}{0.45\textwidth}
 \centering
\includegraphics[width=\textwidth]{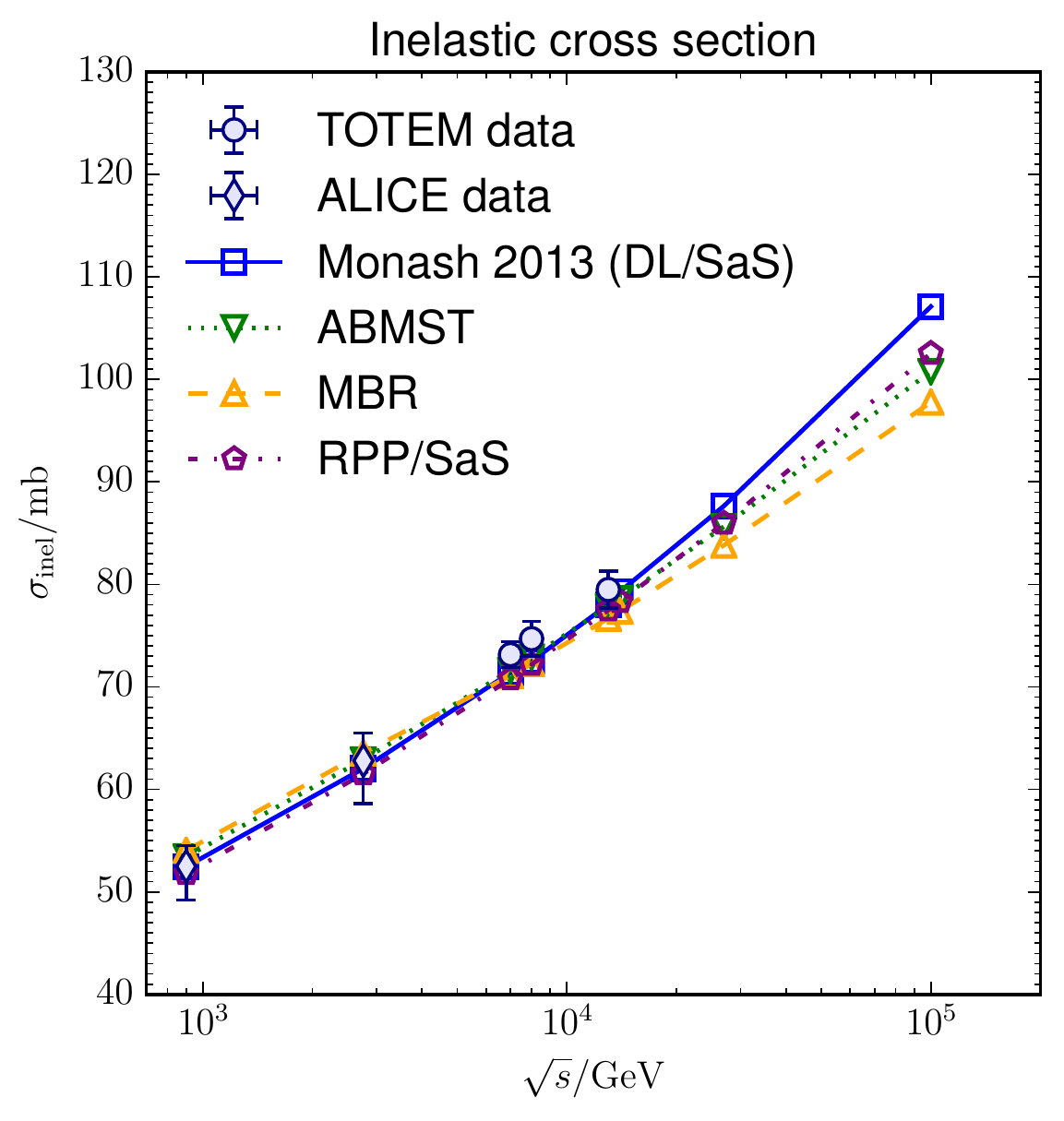}
\caption{}
\label{fig:sigma_inel}
 \end{subfigure}
\caption{Plots showing the energy scaling of \subref{fig:sigma_total} total and \subref{fig:sigma_inel} inelastic cross sections.
Shown for comparison are data collected by the
TOTEM and ALICE collaborations \cite{PhysRevLett.111.012001,0295-5075-101-2-21002,Antchev:2296409,Abelev:2012sea}.}
 \label{fig:sigma_total_inel}
\end{figure}

In \cref{fig:sigma_total_inel}, we show extrapolations of the total and inelastic cross sections predicted  by the selection of parameterisations described in 
\cref{sec:sigmavartune}, compared to measurements by TOTEM and ALICE \cite{PhysRevLett.111.012001,0295-5075-101-2-21002,Antchev:2296409,Abelev:2012sea} at LHC energies. (See \cref{app:tables} for tabulated values.) 
The default parameterisations used in Monash 2013 are in good agreement with the measurements of the total inelastic cross section but are in conflict with the total cross section; this is due to the elastic cross section being too small (by about 10mb). The alternative models were selected for their good agreement with both the total and inelastic cross sections. We note also that the largest inelastic cross section at 100 TeV (corresponding to the highest inelastic event rate), among the models considered here, is predicted by the default parameterisation of the Monash 2013 tune, followed by the RPP/SaS, ABMST, and MBR parameterisations, respectively. 

We now turn to the observables characterising the properties of average inelastic events. For each distribution, each class of tune variants (cross-section parameterisations, PDFs, and CR) is represented as a separate coloured band. The central predictions are also shown, and correspond to the (default) Monash 2013 tune of \textsc{Pythia} 8.2.
Tabulated values for both the central predictions and the associated uncertainties, calculated by taking the minimum and maximum values across all variations in the envelope at each energy,
are given in the appendix in  \cref{table:NandPT,table:etaSlices}.

\begin{figure}[tp]
 \centering
\begin{subfigure}{0.48\textwidth}
 \centering
\includegraphics[width=\textwidth]{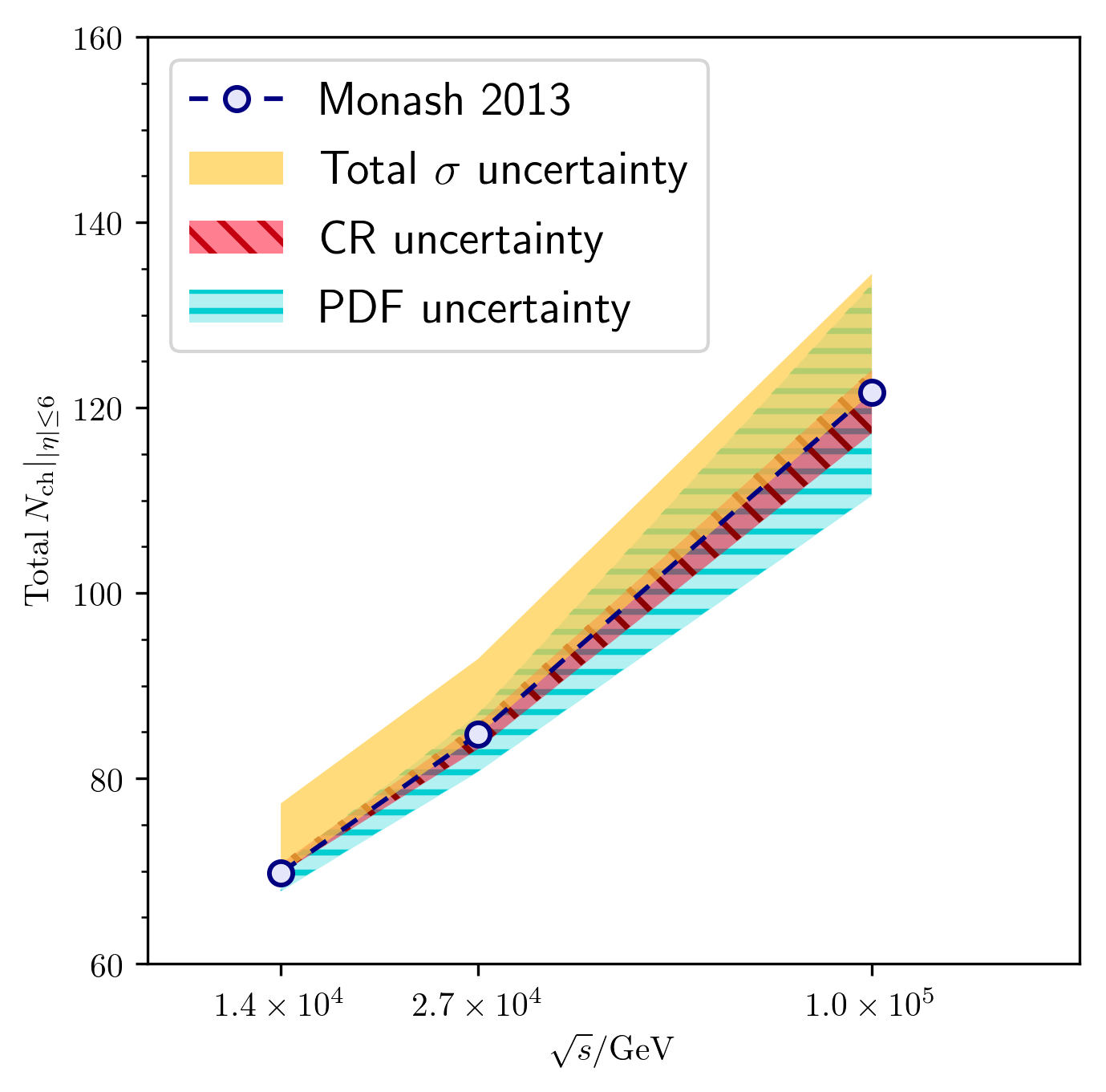}
\caption{}
\label{fig:Nch}
 \end{subfigure}
\begin{subfigure}{0.48\textwidth}
 \centering
\includegraphics[width=\textwidth]{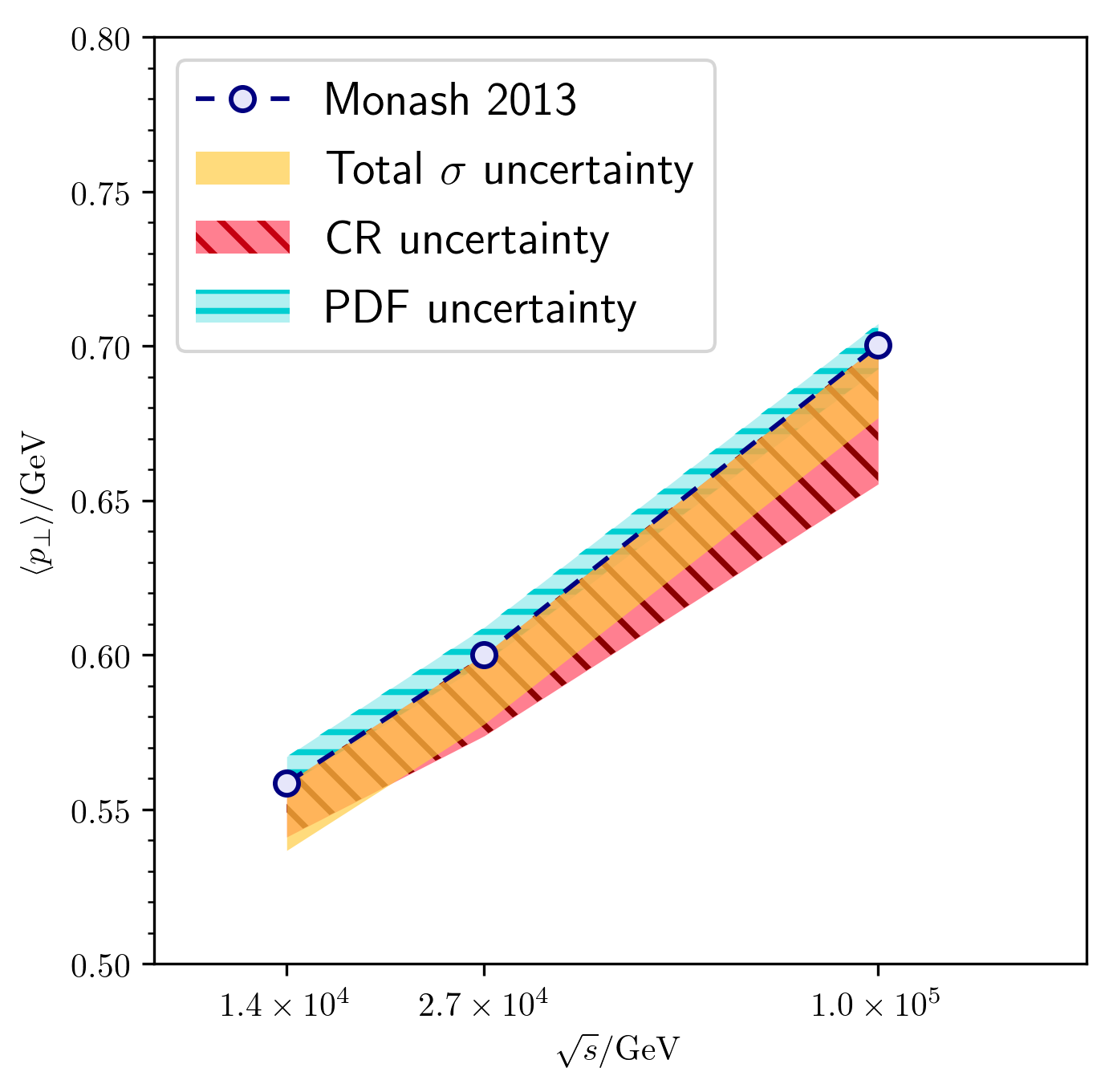}
\caption{}
\label{fig:avePT}
 \end{subfigure}
\caption{Energy scaling of \subref{fig:Nch} the total number of charged tracks inside $|\eta|=6$, and \subref{fig:avePT} the average transverse momentum of charged tracks.}
 \label{fig:NandPT}
\end{figure}

\begin{figure}[tp]
 \centering
\includegraphics[width=0.48\textwidth]{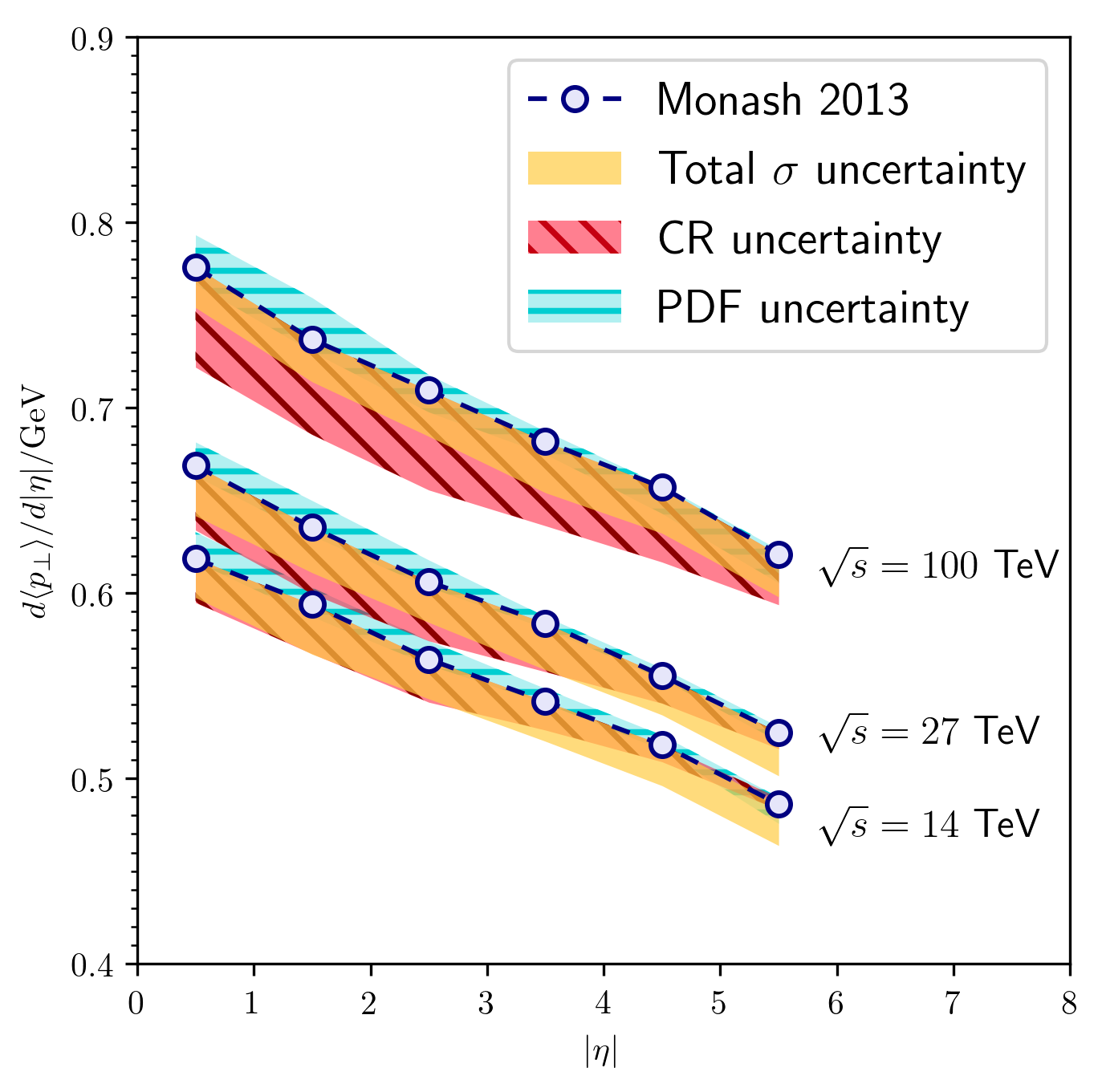}
\caption{Differential average transverse momentum of charged tracks as a function of pseudorapidity.}
 \label{fig:davePTdeta}
\end{figure}

\begin{figure}[tp]
 \centering
\includegraphics[width=0.9\textwidth]{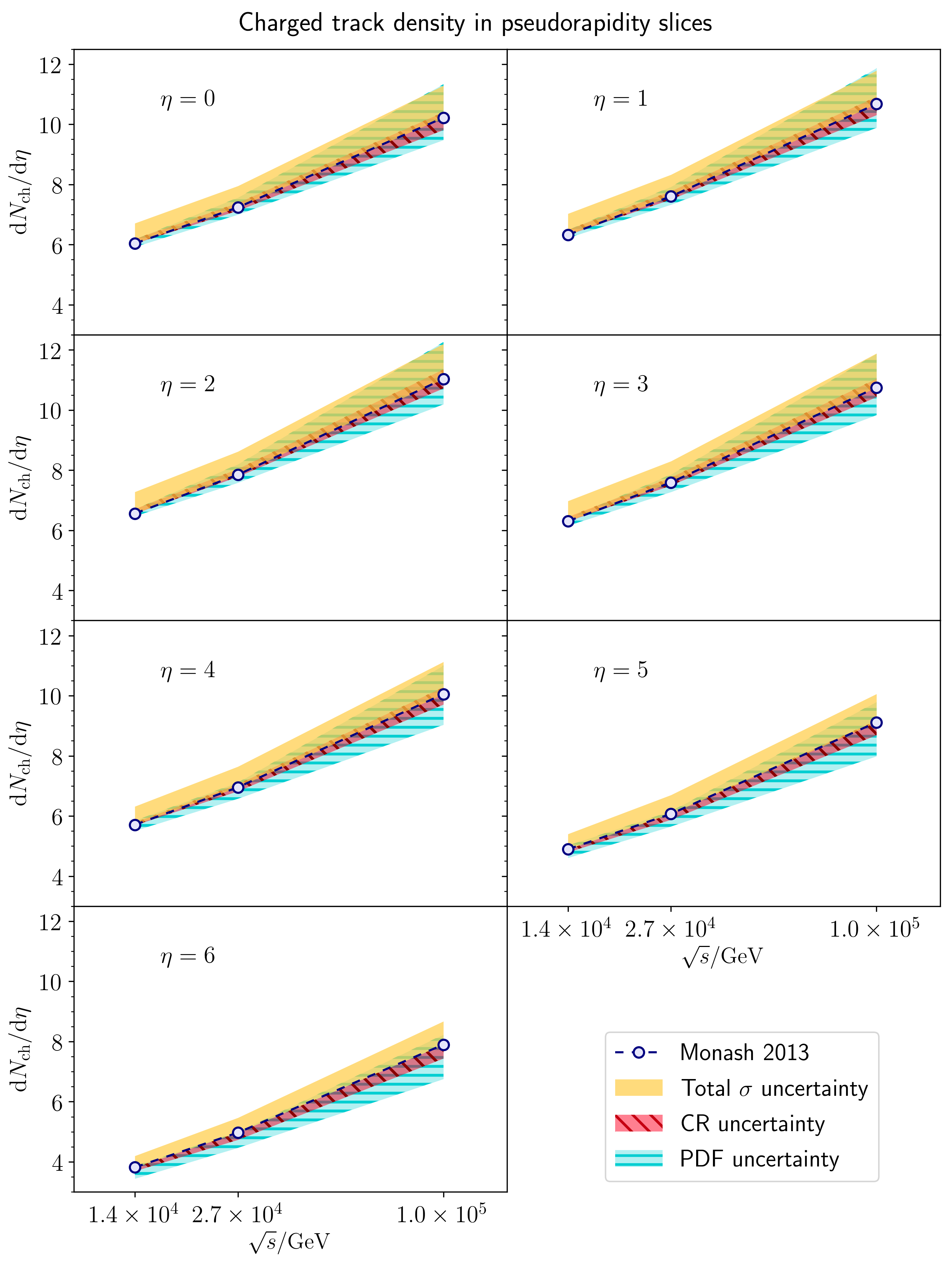}
\caption{Plots showing the scaling of the charged track density in slices of $\eta$  as a function of collider CM energy $\sqrt{s}$.}
 \label{fig:dNchdeta}
\end{figure}

\begin{figure}[tp]
 \centering
\includegraphics[width=0.9\textwidth]{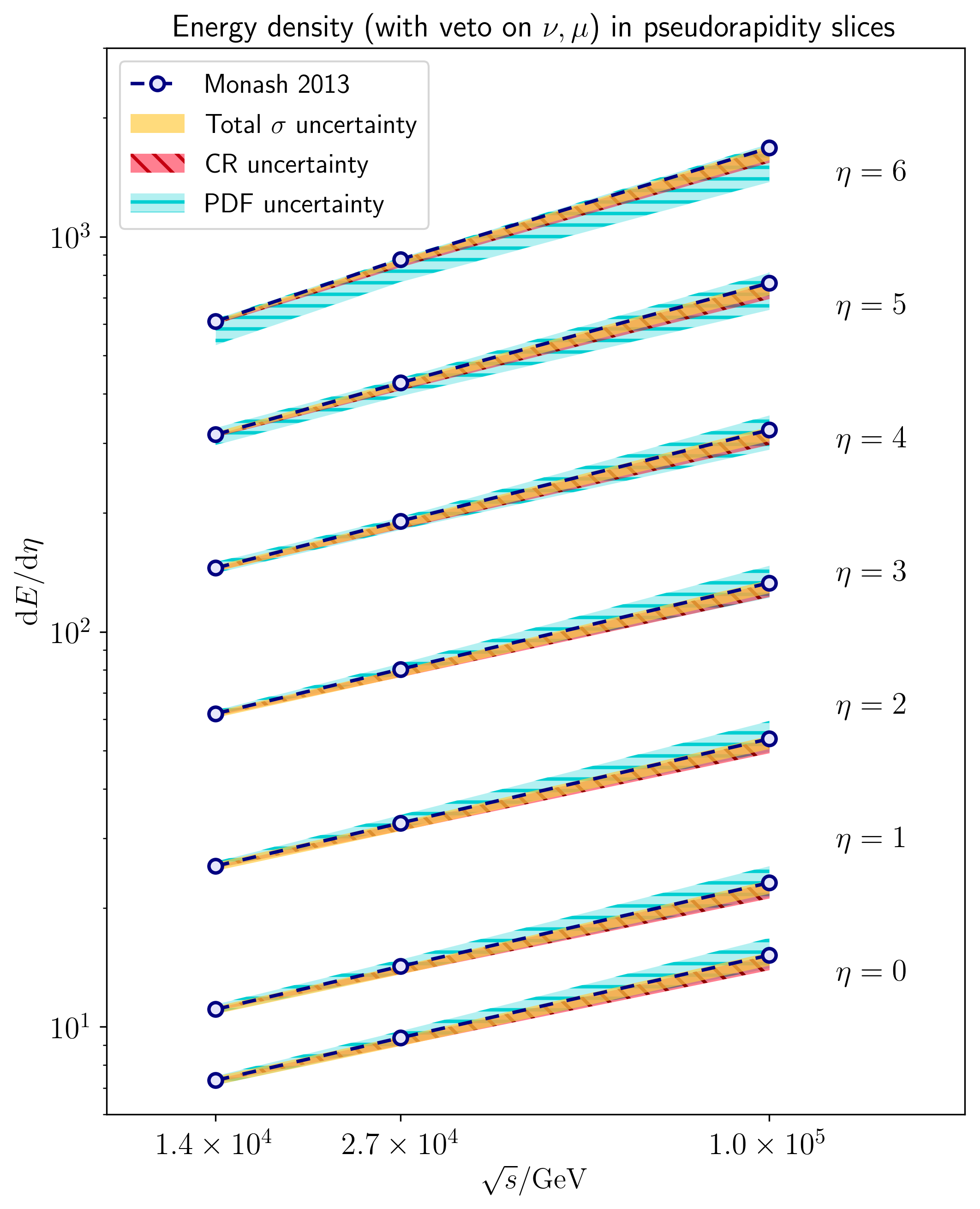}
\caption{Plots showing the scaling of the energy density of charged tracks in slices of $\eta$ as a function of collider CM energy $\sqrt{s}$.}
 \label{fig:dEdeta_veto}
\end{figure}

Below, in figs.~\ref{fig:NandPT} -- \ref{fig:dEdeta_veto}, we show predictions for the following quantities:
\begin{itemize}
 \item The total number of charged tracks inside $|\eta|=6.0$, shown in \cref{fig:Nch}.
 \item The average transverse momentum of charged tracks, shown in \cref{fig:avePT}. Also shown, in \cref{fig:davePTdeta}, is the differential  average transverse momentum of charged tracks as a function of absolute pseudorapidity.
 \item The charged track density in slices of pseudorapidity, \cref{fig:dNchdeta}.
 \item The energy density of all final state particles  in slices of pseudorapidity, \cref{fig:dEdeta_veto}.
\end{itemize}
The events counted in distributions of charged tracks (unless specified otherwise) were required to have at least one charged track inside $|\eta|=6.5$.
Furthermore events were generated such that only those particles having proper lifetimes $c\tau_0 < 10$ mm are decayed (by setting:
\texttt{ParticleDecays:limitTau0 = On} and \texttt{ParticleDecays:tau0Max = 10.0}). In the distribution of energy density, all final state particles \textit{excluding} neutrinos and muons
were counted. As a cross check, the fully inclusive distributions were also calculated, but the inclusion of neutrinos and muons only give rise to per-mille level differences. 

We note that our benchmark predictions for the Monash tune of \textsc{Pythia} are consistent\footnote{In \cite{dEnterria:2016oxo} there were some inconsistencies in the setup for the predictions of $\langle p_T \rangle$ 
between that described in the main text and that shown on the plot; we find agreement if we use the setup shown on the plot.} with the earlier study of \cite{dEnterria:2016oxo}. We emphasise that although \cite{dEnterria:2016oxo}
considered a similar set of observables to ours, they were mainly concerned with the differences between a range of qualitatively different generators, while our study primarily addresses how the modelling uncertainties for \textsc{Pythia} scale with energy.

So far, we have focused on the  modelling  of  nonperturbative  QCD  effects as the dominant source of uncertainty on observables sensitive to the low-scale physics which dominates the bulk of inelastic events. By contrast, uncertainties for hard infrared-safe observables are mainly perturbative in origin, and are usually typified by performing variations of the renormalisation and factorisation scales. As an example, we here provide predictions for the inclusive jet cross sections at next-to-leading order in QCD for ${p_\perp}_j \geq 50$ GeV and ${p_\perp}_j \geq 100$ GeV using \texttt{MG5\_aMC@NLO}\cite{Alwall:2014hca}
matched to \textsc{Pythia}. Jets are defined using the anti-$k_T$ jet-finding algorithm with $R=0.4$. 
Scale variations are performed by varying $\mu_R$ and $\mu_F$ independently by between factors as 2.0 and 0.5\footnote{Scale variations were performed using the conventional method of 7-point variation.}.
The central scale choice used is the default in \texttt{MG5\_aMC@NLO}, namely half the sum of transverse masses over all particles $\frac{\sum m_{T}}{2}$ (equivalent to $H_T/2$ for massless particles).  

In addition we provide predictions for two choices of PDF set, namely NNPDF 2.3 (which is the default both for \texttt{MG5\_aMC@NLO} and the Monash 2013 tune) and HERAPDF 1.5; these were selected as having the largest shape differences over a range of $x$ values. Note that in \texttt{MG5\_aMC@NLO} we use the NLO set for these PDFs, while in the shower and generation of MPI in \textsc{Pythia} we use the LO set so as not to break the underlying-event tuning\footnote{We note that it is appropriate to use LO PDFs in the shower even for cross sections matched to NLO since the splitting kernels used in the shower are still LO; for a longer discussion on this matter see: \url{http://home.thep.lu.se/~torbjorn/pdfdoc/pdfwarning.pdf}. }.
  
\begin{figure}[tp]
 \centering
\begin{subfigure}{0.45\textwidth}
 \centering
\includegraphics[width=\textwidth]{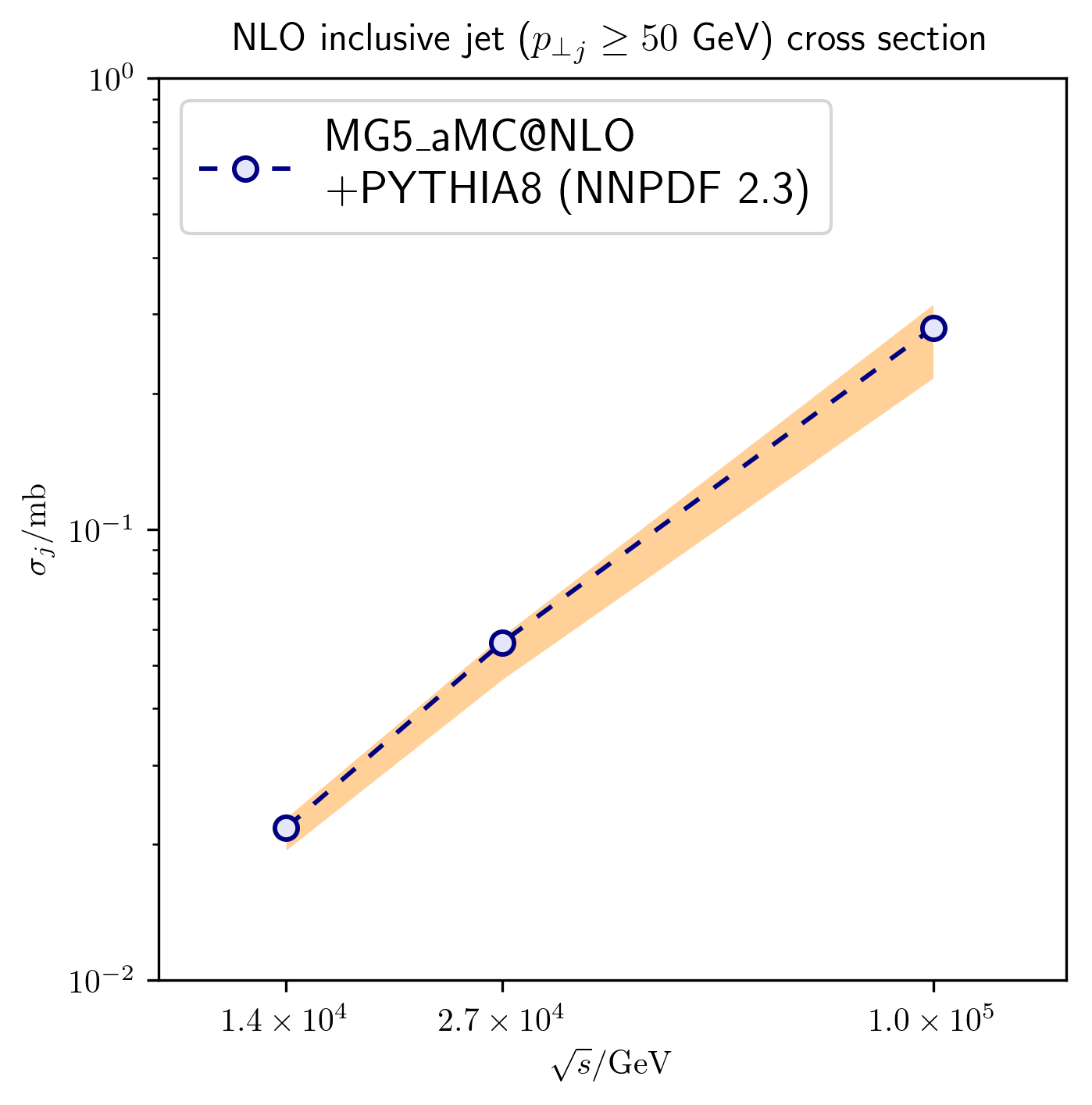}
\caption{}
\label{fig:jet50}
 \end{subfigure}
\begin{subfigure}{0.45\textwidth}
 \centering
\includegraphics[width=\textwidth]{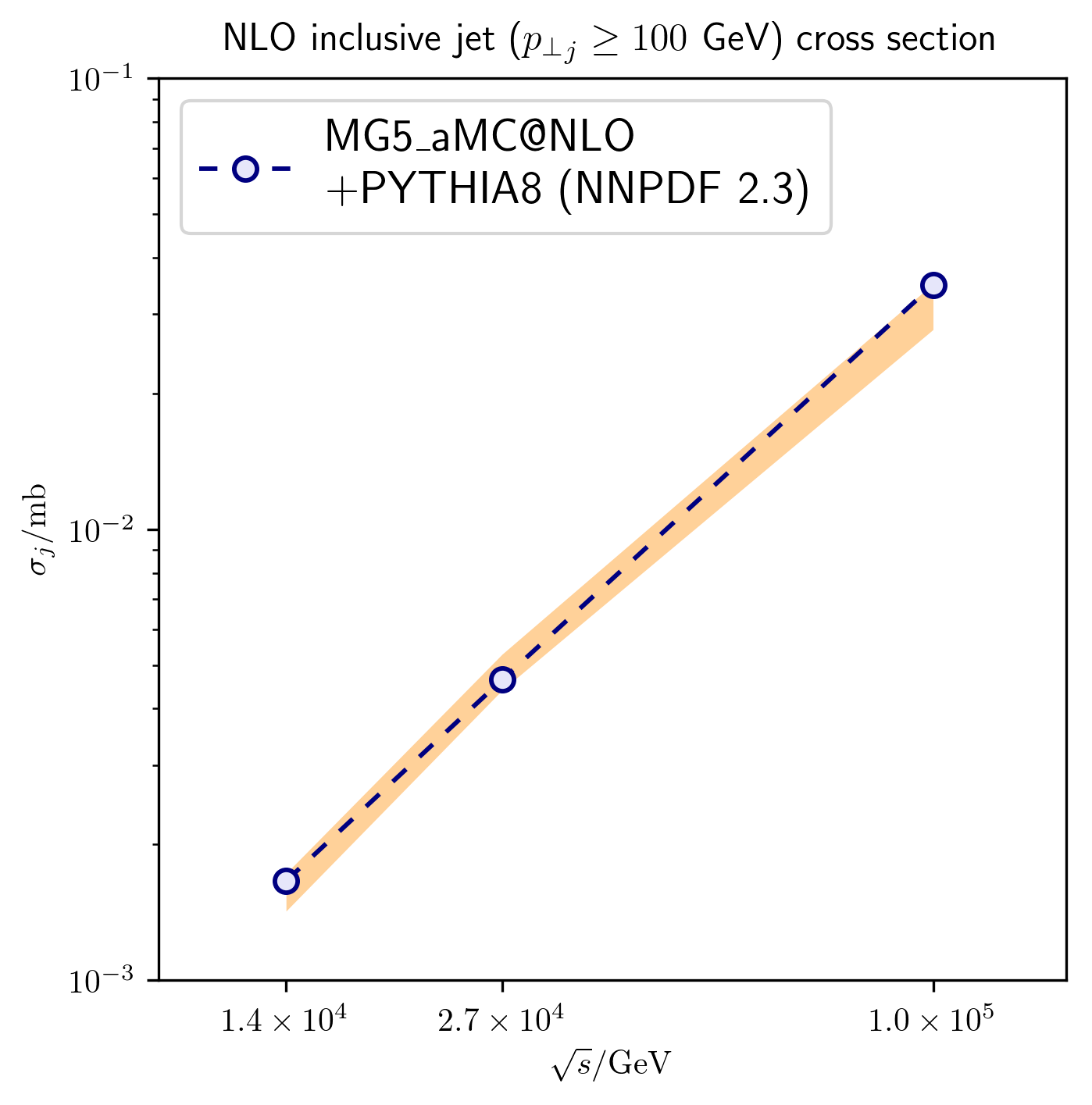}
\caption{}
\label{fig:jet100}
 \end{subfigure}
 \begin{subfigure}{0.45\textwidth}
 \centering
\includegraphics[width=\textwidth]{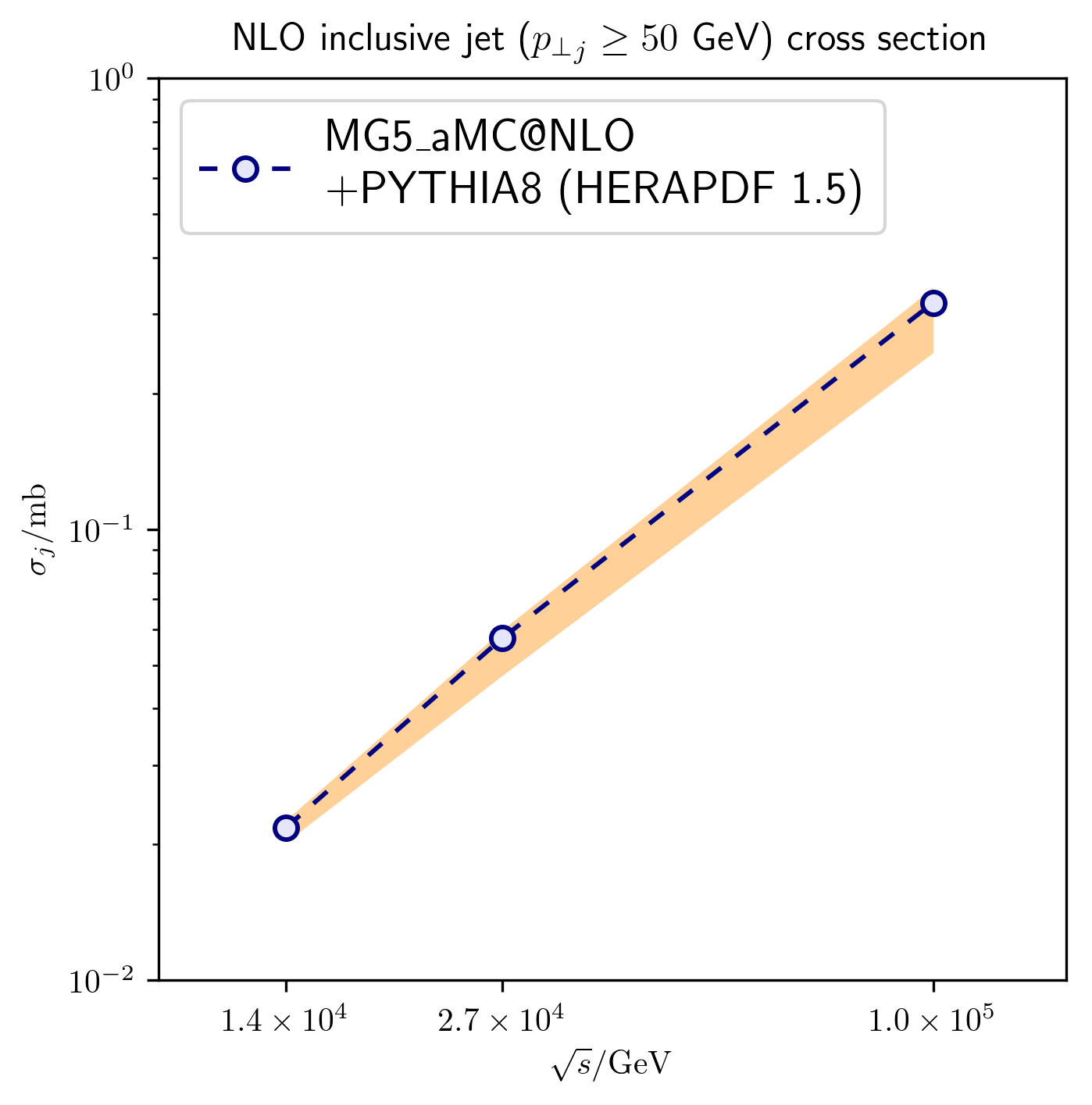}
\caption{}
\label{fig:jet50hera}
 \end{subfigure}
\begin{subfigure}{0.45\textwidth}
 \centering
\includegraphics[width=\textwidth]{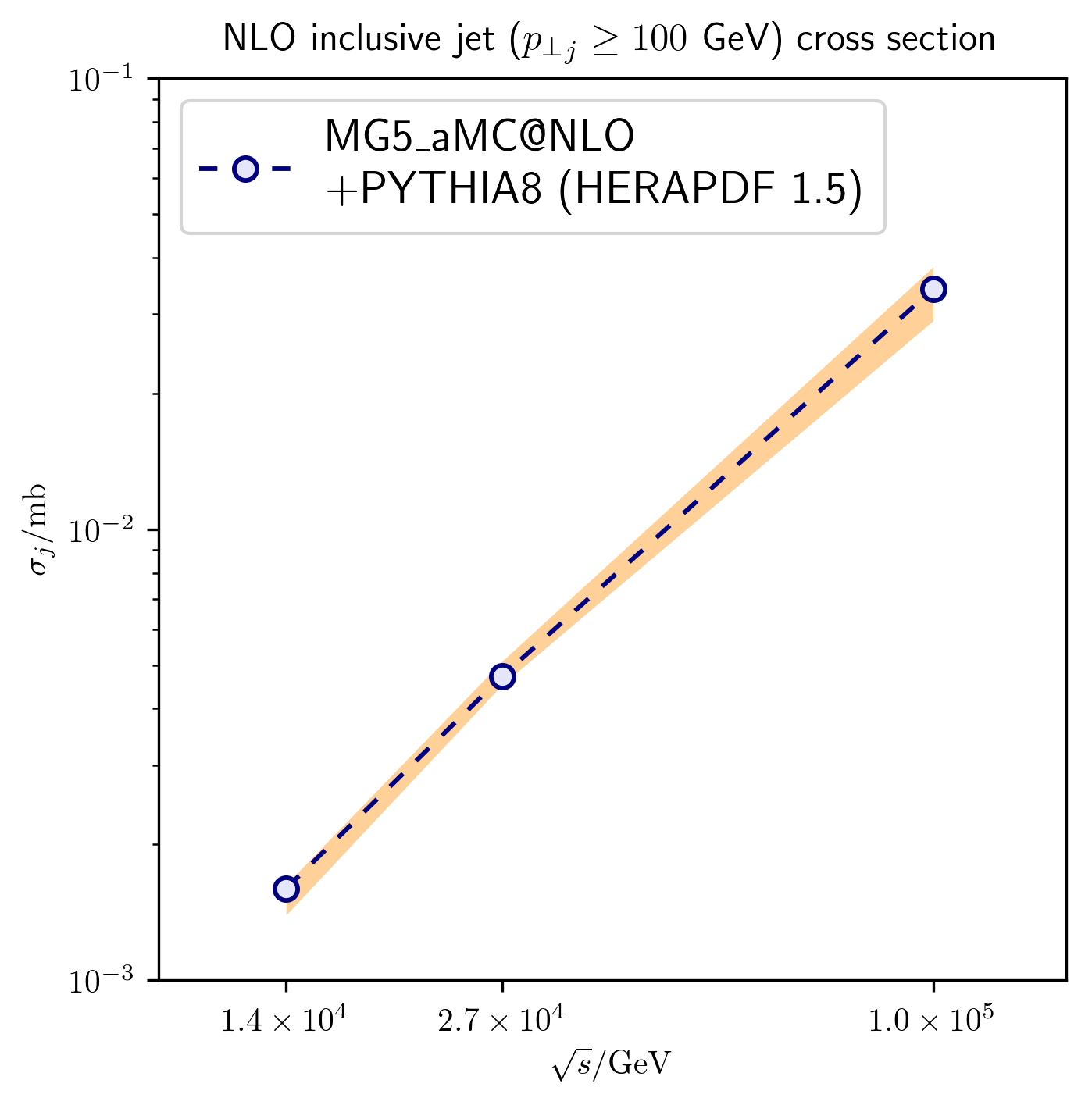}
\caption{}
\label{fig:jet100hera}
 \end{subfigure}
\caption{Plots showing the energy scaling of NLO inclusive jet cross sections, defined as having at least one jet with $p_T>50$ GeV in \subref{fig:jet50}, \subref{fig:jet50hera}, and $p_T>100$ GeV in \subref{fig:jet100}, \subref{fig:jet100hera} using the anti-$k_T$ algorithm with $R=0.4$. The shaded bands denote scale variations with respect to $\mu_R$ and $\mu_F$.
\subref{fig:jet50}, \subref{fig:jet100} correspond to the PDF set NNPDF 2.3 , while \subref{fig:jet50hera},\subref{fig:jet100hera} use the set HERAPDF 1.5.}
 \label{fig:sigma_jet}
\end{figure}
The results are shown in \cref{fig:sigma_jet}, for fully hadronised final-state particles. (See \cref{app:tables} for tabulated values.) 
For the modelling of non-perturbative effects we select the tune appropriate to the LO PDFs\footnote{Had we chosen to use NLO PDFs inside the shower it would have been necessary to re-tune for these choices.}, with the default settings for NNPDF 2.3, and the modifications shown in the fourth column of \cref{table:PDFtunes} for HERAPDF 1.5.  In the results for HERAPDF we considered both the upper and lower variation
for the parameter \texttt{MultipartonInteractions:ecmPow}; the difference was found to be at most 1\% for  $p_T>50$ GeV at 100 TeV and otherwise considerably less. As this is notably less than the size of scale uncertainties, we therefore show only the results for the lower variation.
The difference between the two PDF sets is somewhat larger, but nevertheless much smaller than scale uncertainties. 

As an explicit check that the inclusive cross sections are indeed relatively insensitive to non-perturbative uncertainties, we note that the extreme variation obtained by switching off hadronisation and colour-reconnection in \textsc{Pythia} only modifies these cross sections by approximately 3-4\% at $\sqrt{s}= 14$ TeV and by 1-3\% at $\sqrt{s}= 100$ TeV. 

We note that the earlier study of \cite{Bothmann:2016loj} also made predictions for inclusive jet cross sections for a 100 TeV $p\bar{p}$ collider; however, this focused primarily on the scaling of jet rates and jet substructure (and did not include NLO predictions for the pure QCD single inclusive jet cross section). In addition, \cite{Bothmann:2016loj} considered several different hard processes and their impact for BSM searches. 

\FloatBarrier

\section{Summary and Conclusions \label{sec:conclusions}}

We have presented predictions for the expectation values of some basic observables characterising inelastic event properties at future hadron colliders, focusing on CM energies of 14, 27, and 100 TeV. The predictions are based on extrapolations made with the Monash 2013 tune of the \textsc{Pythia} 8.2 event generator as the baseline and uncertainties were estimated by variations of its PDFs, cross-section parametrisations, colour-reconnection modelling, and MPI $p_{\perp 0}$ regularisation scale, subject to constraints imposed by current LHC measurements at $\sqrt{s} = 0.9$--$13$ TeV. 

The total proton-proton cross section is expected to grow from about 108 mb at  $\sqrt{s} =14$ TeV to about 145 mb at  $\sqrt{s} =100$ TeV, though we note that the 1992 Donnachie-Landshoff fit used by the baseline Monash tune underestimates these numbers by about 5\% mainly due to its elastic cross section being too small and already in conflict with LHC data. We therefore \emph{do not} advise to use the Monash 2013 extrapolations for total and elastic cross sections. 

The inelastic proton-proton cross section is expected to rise from 77--80 mb at $\sqrt{s}=14$ TeV to 83--88 mb at $\sqrt{s}=27$ TeV, to 97--108 mb at $\sqrt{s}=100$ TeV, with the Monash 2013 predictions being the highest among the parametrisations considered in this study and hence relatively conservative  within the context of this study. 

The total number of charged particles inside $|\eta|<6$ (with $c\tau_0 \ge 10\ \mathrm{mm}$) per inelastic event grows by slightly less than a factor of 2 --- from about 70 at $\sqrt{s}=14$ TeV to about 120 at $\sqrt{s}=100$ TeV, with estimated uncertainties of $\sim 10\%$. The average $p_\perp$ of those particles also increases, albeit only slightly, starting from $\left<p_\perp\right> \sim 460$ MeV at $\sqrt{s}=14$ TeV and increasing by 50 -- 100 MeV when extrapolating to $\sqrt{s}=100$ TeV.

In the central region of the detector, the amount of energy deposited per inelastic event grows by about a factor of 2 --- from $\sim$ 7 GeV per unit $\eta$ at $\sqrt{s}=14$ TeV 
to $\sim 15$ GeV per unit $\eta$ at $\sqrt{s}=100$ TeV. 
At high rapidities, much closer to the beam, the total amount of energy deposited is of course much larger, and it is also predicted to scale faster. At $|\eta| = 6$, we estimate about 600 GeV of total energy per unit rapidity at 14 TeV, while we predict $\sim$ 1700 GeV per unit rapidity at 100 TeV. For the highest rapidities, it is also worth remarking that our choice of reference model, the Monash 2013 tune, lies towards the upper limit of the range spanned by our uncertainty estimates, cf.~\cref{fig:dEdeta_veto}, hence its predictions of the maximum energy densities that would be faced by detectors in this region can be considered relatively conservative at least within the context of the variations we have studied. 

The cross section for anti-$k_T$ jets with $\Delta R=0.4$ and 
$p_\perp \ge 50$ GeV is predicted to increase faster than the relative increase in CM energy, by a factor of $\sim 13$ from $\sqrt{s}=14$ TeV to $\sqrt{s}=100$ TeV, while the cross section for a cut of $p_\perp \ge 100$ GeV is expected to increase by about a factor of $21$ over the same range.  

We hope this investigation, and the LHC-vetted \textsc{Pythia} parameter settings that we have developed in the course of it, will prove useful to the exploration of phenomenology and detector design concepts for future hadron colliders. 

\subsection*{Acknowledgements}  HB is funded by the Australian Research Council via Discovery Project 
DP170100708 --- ``Emergent Phenomena in Quantum Chromodynamics''. PS is the recipient of an Australian Research Council Future Fellowship FT1310744 --- ``Virtual Colliders''. Work also supported in part by 
the European Union’s Horizon 2020 research and innovation programme under the 
Marie Sk\l{}odowska-Curie grant agreement No 722105 --- ``MCnetITN3''. 

\bibliography{refs}
\bibliographystyle{JHEP}  

\clearpage
\appendix
\section{Tables}
\label{app:tables}

\begin{table}[h!t]
\centering
\captionsetup{width=1.0\textwidth}
\begin{tabular}{|c|c|c|c|c|}
\hline
\multirow{2}{*}{$\sqrt{s}/\mathrm{TeV}$} & \multicolumn{4}{|c|}{$\sigma_\mathrm{total}/\mathrm{mb}$} \\
\hhline{~----}
 & Monash 2013 (DL/SaS) & ABMST & MBR & RPP/SaS \\
\hline
14 & 101.51 & 108.25 & 109.49 & 107.85\\
27 & 112.87 & 119.81 & 121.28 & 120.20\\
100 & 139.46 & 144.71 & 148.05 & 147.54\\
\hline
\end{tabular}
\caption{Table comparing predictions for the total cross section at $\sqrt{s}$ = 14,27,100 TeV.}
\label{table:sigma_total}
\end{table}

\begin{table}[h!t]
\centering
\captionsetup{width=1.0\textwidth}
\begin{tabular}{|c|c|c|c|c|}
\hline
\multirow{2}{*}{$\sqrt{s}/\mathrm{TeV}$} & \multicolumn{4}{|c|}{$\sigma_\mathrm{inel}/\mathrm{mb}$} \\
\hhline{~----}
 & Monash 2013 (DL/SaS) & ABMST & MBR & RPP/SaS \\
\hline
14 & 79.31 & 78.62 & 77.39 & 78.33\\
27 & 87.65 & 85.64 & 83.82 & 85.96\\
100 & 107.10 & 100.77 & 97.77 & 102.52\\
\hline
\end{tabular}
\caption{Table comparing predictions for the inelastic cross section at $\sqrt{s}$ = 14,27,100 TeV.}
\label{table:sigma_inel}
\end{table}

\begin{table}[h!tp]
\centering
\captionsetup{width=1.0\textwidth}
\begin{tabular}{|c|c|}
\hline
$\sqrt{s}/\mathrm{TeV}$ & $\sigma,{p_\bot}_j \geq 50 \mathrm{GeV} /\mu\mathrm{b}$ \\
\hline
14 & $21.77^{+5.2\mathrm{\%}}_{-10.8\mathrm{\%}}$\\
27 & $56.12^{+3.3\mathrm{\%}}_{-17.3\mathrm{\%}}$\\
100 & $279.44^{+12.7\mathrm{\%}}_{-22.6\mathrm{\%}}$\\
\hline
\end{tabular}\hspace*{1cm}\begin{tabular}{|c|c|}
\hline
$\sqrt{s}/\mathrm{TeV}$ & $\sigma,{p_\bot}_j \geq 100 \mathrm{GeV} /\mu\mathrm{b}$ \\
\hline
14 & $1.66^{+3.9\mathrm{\%}}_{-14.2\mathrm{\%}}$\\
27 & $4.64^{+14.0\mathrm{\%}}_{-5.1\mathrm{\%}}$\\
100 & $34.82^{+0.5\mathrm{\%}}_{-20.4\mathrm{\%}}$\\
\hline
\end{tabular}
\caption{Table showing the NLO inclusive jet cross section for (left) ${p_\bot}_j \geq 50$ GeV and (right) ${p_\bot}_j \geq 100$ GeV, as predicted by \texttt{MG5\_aMC@NLO} + \textsc{Pythia} 8.2 for $\sqrt{s}$ = 14,27,100 TeV using the PDF set NNPDF 2.3. Jets are found using the anti-$k_T$ jet algorithm with $R=0.4$. Uncertainties are shown as percentages and are found as described in the text.}
\label{table:sigma_jet}
\end{table}

\begin{table}[h!tp]
\centering
\captionsetup{width=1.0\textwidth}
\begin{tabular}{|c|c|}
\hline
$\sqrt{s}/\mathrm{TeV}$ & $\sigma,{p_\bot}_j \geq 50 \mathrm{GeV} /\mu\mathrm{b}$ \\
\hline
14 & $21.80^{+3.9\mathrm{\%}}_{-7.3\mathrm{\%}}$\\
27 & $57.32^{+4.3\mathrm{\%}}_{-17.4\mathrm{\%}}$\\
100 & $317.69^{+8.1\mathrm{\%}}_{-22.5\mathrm{\%}}$\\
\hline
\end{tabular}\hspace*{1cm}\begin{tabular}{|c|c|}
\hline
$\sqrt{s}/\mathrm{TeV}$ & $\sigma,{p_\bot}_j \geq 100 \mathrm{GeV} /\mu\mathrm{b}$ \\
\hline
14 & $1.60^{+2.1\mathrm{\%}}_{-12.6\mathrm{\%}}$\\
27 & $4.72^{+8.0\mathrm{\%}}_{-4.2\mathrm{\%}}$\\
100 & $34.09^{+11.9\mathrm{\%}}_{-15.0\mathrm{\%}}$\\
\hline
\end{tabular}
\caption{Table showing the NLO inclusive jet cross section for (left) ${p_\bot}_j \geq 50$ GeV and (right) ${p_\bot}_j \geq 100$ GeV,as predicted by \texttt{MG5\_aMC@NLO} + \textsc{Pythia} 8.2 for $\sqrt{s}$ = 14,27,100 TeV using the PDF set HERAPDF 1.5. Jets are found using the anti-$k_T$ jet algorithm with $R=0.4$. Uncertainties are shown as percentages and are found as described in the text.}
\label{table:sigma_jet_hera}
\end{table}

\begin{table}[h!t]
\centering
\captionsetup{width=1.0\textwidth}
\begin{tabular}{|c|c|}
\hline
$\sqrt{s}/\mathrm{TeV}$ & Total $N_{ch}, |\eta|\leq 6$ \\
\hline
14 & $69.78^{+10.8\mathrm{\%}}_{-2.8\mathrm{\%}}$\\
27 & $84.74^{+9.7\mathrm{\%}}_{-4.7\mathrm{\%}}$\\
100 & $121.68^{+10.5\mathrm{\%}}_{-9.1\mathrm{\%}}$\\
\hline
\end{tabular}\hspace*{1cm}\begin{tabular}{|c|c|}
\hline
$\sqrt{s}/\mathrm{TeV}$ & $\langle p_\bot \rangle/\mathrm{GeV}$ \\
\hline
14 & $0.56^{+1.8\mathrm{\%}}_{-4.3\mathrm{\%}}$\\
27 & $0.59^{+1.4\mathrm{\%}}_{-4.9\mathrm{\%}}$\\
100 & $0.71^{+0.3\mathrm{\%}}_{-8.0\mathrm{\%}}$\\
\hline
\end{tabular}
\caption{Table showing (left) the total number of charged tracks inside $|\eta|=6$, and (right) the average transverse momentum of charged tracks, as predicted by the Monash 2013 tune of \textsc{Pythia} 8.2, at $\sqrt{s}$ = 14,27,100 TeV. Only events containing at least 1 charged track inside $|\eta|<6.5$ are counted. Only particles with proper lifetimes $c \tau_0 < 10$ mm are considered to decay. Uncertainties are shown as percentages and are found as described in the text.}
\label{table:NandPT}
\end{table}

\begin{sidewaystable}[h!t]
\centering
\captionsetup{width=1.0\textwidth}
\begin{tabular}{|c|c|c|c|c|c|c|c|}
\hline
\multirow{2}{*}{$\sqrt{s}/\mathrm{TeV}$} & \multicolumn{7}{|c|}{$\mathrm{d}N_\mathrm{ch}/\mathrm{d}\eta$} \\
\hhline{~-------}
 & $\eta=0$ & $\eta=1$ & $\eta=2$ & $\eta=3$ & $\eta=4$ & $\eta=5$ & $\eta=6$ \\
\hline
14 & $6.04^{+11.1\mathrm{\%}}_{-1.9\mathrm{\%}}$ & $6.33^{+11.1\mathrm{\%}}_{-1.6\mathrm{\%}}$ & $6.56^{+11.0\mathrm{\%}}_{-1.4\mathrm{\%}}$ & $6.30^{+10.7\mathrm{\%}}_{-2.4\mathrm{\%}}$ & $5.71^{+10.7\mathrm{\%}}_{-3.5\mathrm{\%}}$ & $4.89^{+10.3\mathrm{\%}}_{-5.8\mathrm{\%}}$ & $3.82^{+9.9\mathrm{\%}}_{-9.9\mathrm{\%}}$\\
27 & $7.24^{+9.8\mathrm{\%}}_{-3.3\mathrm{\%}}$ & $7.61^{+9.5\mathrm{\%}}_{-3.6\mathrm{\%}}$ & $7.85^{+9.8\mathrm{\%}}_{-3.4\mathrm{\%}}$ & $7.59^{+9.3\mathrm{\%}}_{-4.2\mathrm{\%}}$ & $6.96^{+9.9\mathrm{\%}}_{-5.4\mathrm{\%}}$ & $6.07^{+10.3\mathrm{\%}}_{-6.9\mathrm{\%}}$ & $4.97^{+9.9\mathrm{\%}}_{-10.2\mathrm{\%}}$\\
100 & $10.22^{+10.9\mathrm{\%}}_{-7.2\mathrm{\%}}$ & $10.67^{+11.2\mathrm{\%}}_{-7.4\mathrm{\%}}$ & $11.03^{+11.1\mathrm{\%}}_{-7.6\mathrm{\%}}$ & $10.75^{+10.5\mathrm{\%}}_{-8.5\mathrm{\%}}$ & $10.05^{+10.7\mathrm{\%}}_{-10.1\mathrm{\%}}$ & $9.12^{+10.3\mathrm{\%}}_{-12.4\mathrm{\%}}$ & $7.90^{+9.8\mathrm{\%}}_{-14.5\mathrm{\%}}$\\
\hline
\end{tabular}\\[1cm]
\begin{tabular}{|c|c|c|c|c|c|c|c|}
\hline
\multirow{2}{*}{$\sqrt{s}/\mathrm{TeV}$} & \multicolumn{7}{|c|}{$\mathrm{d}E/\mathrm{d}\eta$} \\
\hhline{~-------}
 & $\eta=0$ & $\eta=1$ & $\eta=2$ & $\eta=3$ & $\eta=4$ & $\eta=5$ & $\eta=6$ \\
\hline
14 & $7.32^{+2.7\mathrm{\%}}_{-2.7\mathrm{\%}}$ & $11.07^{+3.0\mathrm{\%}}_{-2.5\mathrm{\%}}$ & $25.53^{+2.6\mathrm{\%}}_{-2.6\mathrm{\%}}$ & $62.05^{+2.0\mathrm{\%}}_{-2.2\mathrm{\%}}$ & $144.99^{+3.5\mathrm{\%}}_{-2.9\mathrm{\%}}$ & $315.59^{+4.1\mathrm{\%}}_{-5.8\mathrm{\%}}$ & $610.13^{+1.6\mathrm{\%}}_{-12.8\mathrm{\%}}$\\
27 & $9.39^{+4.1\mathrm{\%}}_{-4.6\mathrm{\%}}$ & $14.23^{+4.1\mathrm{\%}}_{-4.4\mathrm{\%}}$ & $32.81^{+4.8\mathrm{\%}}_{-4.5\mathrm{\%}}$ & $80.42^{+3.5\mathrm{\%}}_{-4.4\mathrm{\%}}$ & $190.70^{+2.4\mathrm{\%}}_{-5.0\mathrm{\%}}$ & $426.79^{+2.8\mathrm{\%}}_{-7.3\mathrm{\%}}$ & $876.10^{+1.0\mathrm{\%}}_{-12.3\mathrm{\%}}$\\
100 & $15.19^{+10.5\mathrm{\%}}_{-8.3\mathrm{\%}}$ & $23.16^{+10.3\mathrm{\%}}_{-8.7\mathrm{\%}}$ & $53.58^{+11.0\mathrm{\%}}_{-7.9\mathrm{\%}}$ & $132.92^{+10.7\mathrm{\%}}_{-8.2\mathrm{\%}}$ & $324.25^{+8.9\mathrm{\%}}_{-10.8\mathrm{\%}}$ & $763.23^{+6.3\mathrm{\%}}_{-14.4\mathrm{\%}}$ & $1677.63^{+2.6\mathrm{\%}}_{-18.0\mathrm{\%}}$\\
\hline
\end{tabular}
\caption{Tables showing (top) the number density of charged tracks with $c\tau_0 \ge 10$ mm, and (bottom) the energy density of all final-state particles excluding neutrinos and muons, for different $\eta$ slices as predicted by the Monash 2013 tune of \textsc{Pythia} 8.2, at $\sqrt{s}$ = 14,27,100 TeV. Only particles with proper lifetimes $c \tau_0 < 10$ mm are considered to decay. Uncertainties are shown as percentages and are found as described in the text.}
\label{table:etaSlices}
\end{sidewaystable}

\clearpage

\end{document}